\documentclass[journal,draftclsnofoot,onecolumn,12pt]{IEEEtran}

\usepackage{cite}

\ifCLASSINFOpdf
  \usepackage[pdftex]{graphicx}
  \usepackage{epstopdf} 
\else
\fi
\usepackage{amsmath}
\usepackage{algorithm}
\usepackage{algorithmic}

  \usepackage[caption=false,font=footnotesize]{subfig}
\usepackage{url}

\usepackage{amsthm}

\newtheorem{theorem}{Theorem}
\newtheorem{corollary}{Corollary}[theorem]

\theoremstyle{definition}

\newtheoremstyle{myremark}
{}
{}
{}
{0pt}
{\bfseries}
{.}
{ }
{\thmname{#1}\thmnumber{ #2}: \thmnote{#3}}

\theoremstyle{theoremdd}

\theoremstyle{myremark}

\newcommand{\comment}[1]{{}} 

\usepackage{amssymb}
\usepackage{xspace}
\usepackage{bm}

\newcommand{\parens}[1]{{\left(#1\right)}\xspace}
\newcommand{\brackets}[1]{{\left[#1\right]}\xspace}

\newcommand{\bars}[1]{{\left\vert#1\right\vert}\xspace}

\newcommand{\complex}{\ensuremath{\mathbb{C}}\xspace}

\newcommand{\inv}{\ensuremath{^{-1}}\xspace}
\newcommand{\prob}[1]{\ensuremath{\mathbb{P}\brackets{#1}}\xspace}
\newcommand{\ev}[1]{\ensuremath{\mathbb{E}\brackets{#1}}\xspace}
\newcommand{\var}[1]{\ensuremath{\mathbb{V}\brackets{#1}}\xspace}

\newcommand{\distcgauss}[2]{\ensuremath{\mathcal{N}_{\complex}\parens{#1,#2}}\xspace} 

\newcommand{\todB}[1]{\ensuremath{\brackets{#1}_{\mathrm{dB}}}}

\newcommand{\msnr}{\ensuremath{\mathsf{SNR}}\xspace}

\newcommand{\msinr}{\ensuremath{\mathsf{SINR}}\xspace}

\newcommand{\msir}{\ensuremath{\mathsf{SIR}}\xspace}
\newcommand{\minr}{\ensuremath{\mathsf{INR}}\xspace}
\newcommand{\linr}{\ensuremath{\overline{\mathsf{INR}}}\xspace}
\newcommand{\lsnr}{\ensuremath{\overline{\mathsf{SNR}}}\xspace}

\newcommand{\noisestd}{\ensuremath{\sigma_{\mathrm{n}}}\xspace}
\newcommand{\noisevar}{\ensuremath{\noisestd^2}\xspace}

\newcommand{\Ptx}{\ensuremath{P_{\mathrm{tx}}\xspace}}

\newcommand{\Ppl}{\ensuremath{\mathrm{PL}\xspace}}
\newcommand{\powertx}{\ensuremath{P_{\mathrm{tx}}}\xspace}

\newcommand{\powerdes}{\ensuremath{P_{\mathrm{des}}}\xspace}
\newcommand{\powerint}{\ensuremath{P_{\mathrm{int}}}\xspace}

\newcommand{\radiusearth}{R_{\mathrm{E}}}
\newcommand{\numbeams}{N_{\mathrm{B}}}
\newcommand{\distSR}[1]{\mathsf{SR}\parens{#1}}
\newcommand{\distSSR}[1]{\mathsf{SSR}\parens{#1}}

\usepackage{xspace}

\def\E{\mathbb{E}}

\newcommand{\ea}{\stackrel{(\text{a})}{=}}
\newcommand{\eb}{\stackrel{(\text{b})}{=}}

\def\b0{{\mathbf{0}}}

\usepackage{xcolor}

\usepackage{xspace}
\usepackage[acronym,nogroupskip,nonumberlist,nopostdot]{glossaries}
\makeglossaries

\newacronym{snr}{SNR}{signal-to-noise ratio}
\newacronym{sinr}{SINR}{signal-to-interference-plus-noise ratio}
\newacronym{sir}{SIR}{signal-to-interference ratio}
\newacronym{inr}{INR}{interference-to-noise ratio}
\newacronym{pdf}{PDF}{probability distribution function}
\newacronym{cdf}{CDF}{cumulative distribution function}
\newacronym{leo}{LEO}{low-earth orbit}
\newacronym{frf}{FRF}{frequency reuse factor}
\newacronym{los}{LOS}{line-of-sight}
\newacronym{nlos}{NLOS}{non-line-of-sight}
\newacronym{mimo}{MIMO}{multiple-input multiple-output}
\newacronym{sr}{SR}{Shadowed Rician}
\newacronym{ssr}{SSR}{Squared Shadowed Rician}
\newacronym{5g}{5G}{fifth generation}

\newcommand{\snr}{\gls{snr}\xspace}
\newcommand{\sinr}{\gls{sinr}\xspace}
\newcommand{\sir}{\gls{sir}\xspace}
\newcommand{\inr}{\gls{inr}\xspace}
\newcommand{\los}{\gls{los}\xspace}
\newcommand{\nlos}{\gls{nlos}\xspace}
\newcommand{\sr}{\gls{sr}\xspace}
\newcommand{\ssr}{\gls{ssr}\xspace}
\newcommand{\leo}{\gls{leo}\xspace}
\newcommand{\pdf}{\gls{pdf}\xspace}
\newcommand{\cdf}{\gls{cdf}\xspace}

\newcommand{\gpdf}{\gls{pdf}\xspace}

\newcommand{\gcdf}{\gls{cdf}\xspace}

\newcommand{\gsnr}{\gls{snr}\xspace}
\newcommand{\ginr}{\gls{inr}\xspace}
\newcommand{\gsinr}{\gls{sinr}\xspace}
\newcommand{\gsir}{\gls{sir}\xspace}
\newcommand{\gpsnr}{\glspl{snr}\xspace}

\newcommand{\gpsinr}{\glspl{sinr}\xspace}

\newcommand{\secref}[1]{Section~\ref{#1}}

\newcommand{\tabref}[1]{Table~\ref{#1}}
\newcommand{\figref}[1]{\figurename~\ref{#1}}

\newcommand{\thmref}[1]{Theorem~\ref{#1}}
\newcommand{\corref}[1]{Corollary~\ref{#1}}

\usepackage{mathtools}

\begin{document}

%
\title{Downlink Analysis and Evaluation of Multi-Beam LEO Satellite Communication in Shadowed Rician Channels}
%
%
%

\author{%
    Eunsun Kim,~%
	Ian~P.~Roberts,~%
	and Jeffrey~G.~Andrews%
	\thanks{The authors are with the 6G@UT Research Center and the Wireless Networking and Communications Group at the University of Texas at Austin. This work is an extension of our early work \cite{globecom_mine}. 
	Corresponding author: E.~Kim (esunkim@utexas.edu).}
}

\maketitle

\begin{abstract}
The extension of wide area wireless connectivity to \leo satellite communication systems demands a fresh look at the effects of in-orbit base stations, sky-to-ground propagation, and cell planning.
A multi-beam \leo satellite delivers widespread coverage by forming multiple spot beams that tessellate cells over a given region on the surface of the Earth.
In doing so, overlapping spot beams introduce interference when delivering downlink concurrently in the same area using the same frequency spectrum.  
To permit forecasting of communication system performance, we characterize desired and interference signal powers, along with SNR, INR, SIR, and SINR, under the measurement-backed \sr sky-to-ground channel model.
We introduce a minor approximation to the fading order of \sr channels that greatly simplifies the PDF and CDF of these quantities and facilitates statistical analyses of \leo satellite systems such as probability of outage. 
We conclude this paper with an evaluation of multi-beam \leo satellite communication in \sr channels of varying intensity fitted from existing measurements.
Our numerical results highlight the effects satellite elevation angle has on SNR, INR, and SINR, which brings attention to the variability in system state and potential performance as a satellite traverses across the sky along its orbit.
\end{abstract}

%
\IEEEpeerreviewmaketitle

\glsresetall 

\section{Introduction} \label{sec:introduction}

\Gls{leo} satellite communication systems are experiencing a renaissance.
Deployment costs have dropped dramatically due to new launch technology, both enabling and being enabled by ongoing mass deployments such as SpaceX's Starlink \cite{spaceX} and Amazon's Project Kuiper \cite{kuiper}. 
Efforts such as these are slated to deploy constellations comprised of thousands or even tens of thousands of \leo satellites, targeted to deliver high-capacity broadband connectivity to un/under-served communities as well as supplement existing terrestrial wireless services in well-served areas.
A single base station onboard a \leo satellite can deliver broad coverage by tessellating multiple spot beams whose collective footprint may have a diameter on the order of tens or hundreds of kilometers.
The orbiting nature of \leo satellite constellations, along with characteristics of sky-to-ground propagation, poses link-level and network-level challenges unseen in terrestrial cellular networks. 
The success of \leo satellite communication systems and their role in next-generation connectivity will rely on accurately evaluating their potential through practically-sound analysis and simulation.

Multiple antennas onboard a single satellite allows it to form multiple high gain beams, which has shown to be a promising route to satisfy demand for high data rates and provide broad coverage, both in \leo and geostationary satellite systems \cite{spaceX, kuiper, hts1, hts2, hts3}.
Even with highly directional beams, it is often assumed that multi-beam satellite systems operate in an interference-limited regime due to co-channel interference between the main lobes of neighboring spot beams \cite{MUprecancel,LutzErich2016TtTs}.
Less aggressive frequency reuse and strategic beam steering can be used to combat spot beam interference \cite{ LutzErich2016TtTs,InterferenceCottatellucci,precoding,MUprecancel}. 
As a satellite traverses across the sky along its orbit, its observed beam patterns on the ground distort---even when correcting its beams' steering directions along the way.
This is attributed to fact that delivered antenna gain is the projection of the radiation pattern onto the surface of the Earth.
The elevation angle of a satellite relative to a ground user, therefore, can dictate the quality of service it can deliver to that user.
These factors were less of a concern in geostationary satellite systems due to their near static relative positioning, but the fast orbital speeds of \leo satellites magnify the time-varying nature of these effects, since a ground user is in view of a particular satellite for only several minutes at most \cite{leo_architecture, leo_gen2, leo_cell}.

In addition to the orbiting nature of satellites, sky-to-ground propagation also plays a central role in characterizing the performance of \leo satellite systems. 
It has been observed that received signals from a satellite typically have \los and \nlos components and experience seemingly random fluctuations caused by buildings, trees, and even vegetation \cite{loo,lutz,fontan,barts}. 
Among efforts to characterize this \cite{loo,lutz,fontan,barts, newsimple}, the \sr model \cite{newsimple} has been adopted widely in literature \cite{sr_mrc, sr_relay,sr_stgem,sumsquaredshadowrician,closed_sum_ssr}, as it offers a closed-form \pdf and \cdf and aligns well with measurements \cite{newsimple}.  
In this model, \los and \nlos propagation are combined in a Rician fashion, where the magnitude of each component randomly fluctuates. 
With the magnitude of the channel modeled as an \sr random variable, the signal power is a \ssr random variable, whose \gpdf and \gcdf are derived in \cite{sumsquaredshadowrician,closed_sum_ssr}, along with that for the sum of \ssr random variables.
While these closed-form \pdf and \cdf are somewhat simplified from the other satellite channel models \cite{loo, lutz, fontan, barts, sat_ch_coraz, itur_681_10}, they involve infinite power series and special functions and are more complex than those traditionally used in terrestrial networks. 
As a result, it can be difficult to derive analytical results based on the \sr channel model.

Existing work has evaluated multi-beam satellite communication systems \cite{beam_coverage, leo_comparison, tecno_ec, sumsquaredshadowrician, ntn_leo, lutz }
 but do not account for important practical considerations.
Works \cite{beam_coverage, leo_comparison, tecno_ec} ignore interference between spot beams, while \cite{sumsquaredshadowrician} assumes interference between spot beams to be uncorrelated with a desired signal.
Furthermore, \cite{lutz} does not account for the distorted beam shape when a satellite is not directly overhead, instead assuming perfectly circular coverage on the ground regardless of elevation angle.
In addition, realistic shadowing has not been incorporated in \cite{ntn_leo}, which has instead assumed channels to be unfaded \los channels.
All this motivates the need to appropriately evaluate multi-beam \leo systems while accounting for practical factors that play a central role in determining system performance.

\textbf{We present a model for the analyses of multi-beam \leo satellite systems.}
We recognize that the downlink desired and interference signal powers of a multi-beam satellite are fully correlated rather than independent since they travel along the same sky-to-ground channel.
We leverage this to characterize \snr, \inr, \sir, and \sinr of the system under \sr channels, and in doing so, we derive relations between linearly-related \sr and \ssr random variables.
To facilitate this characterization, we show that rounding the fading order of a \sr channel to an integer can remove infinite series from expressions for its \pdf and \cdf, which in turn yields closed-form statistics, such as expectation.
As an added benefit, this simplifies numerical realization by removing the presence of infinite series.

\textbf{We conduct a performance evaluation of multi-beam \leo satellite systems.}
Through simulation, we incorporate the effects of multi-beam interference, elevation angle, \sr channels, and frequency reuse to investigate their impact on \snr, \inr, and \sinr.
To appropriately model a variety of \sr channels, we employ three shadowing levels---light, average, and heavy---whose statistical parameters have been fitted from measurements \cite{loo, channel_measure, newsimple}.
We show that the system can be heavily interference-limited or noise-limited, depending on elevation angle and shadowing conditions, but frequency reuse can be a reliable way to reduce interference at the cost of bandwidth.
Considering that a \leo satellite traverses the sky over the course of a few minutes at most, the system can swing from interference-limited to noise-limited and back to interference-limited in a single overhead pass.
This, along with other results from our performance evaluation, can drive design decisions such as cell planning and handover and motivate a variety of future work.

\section{System Model} \label{sec:system-model}

\begin{figure*}[t!]
	\centering
	\includegraphics[width=1\linewidth]{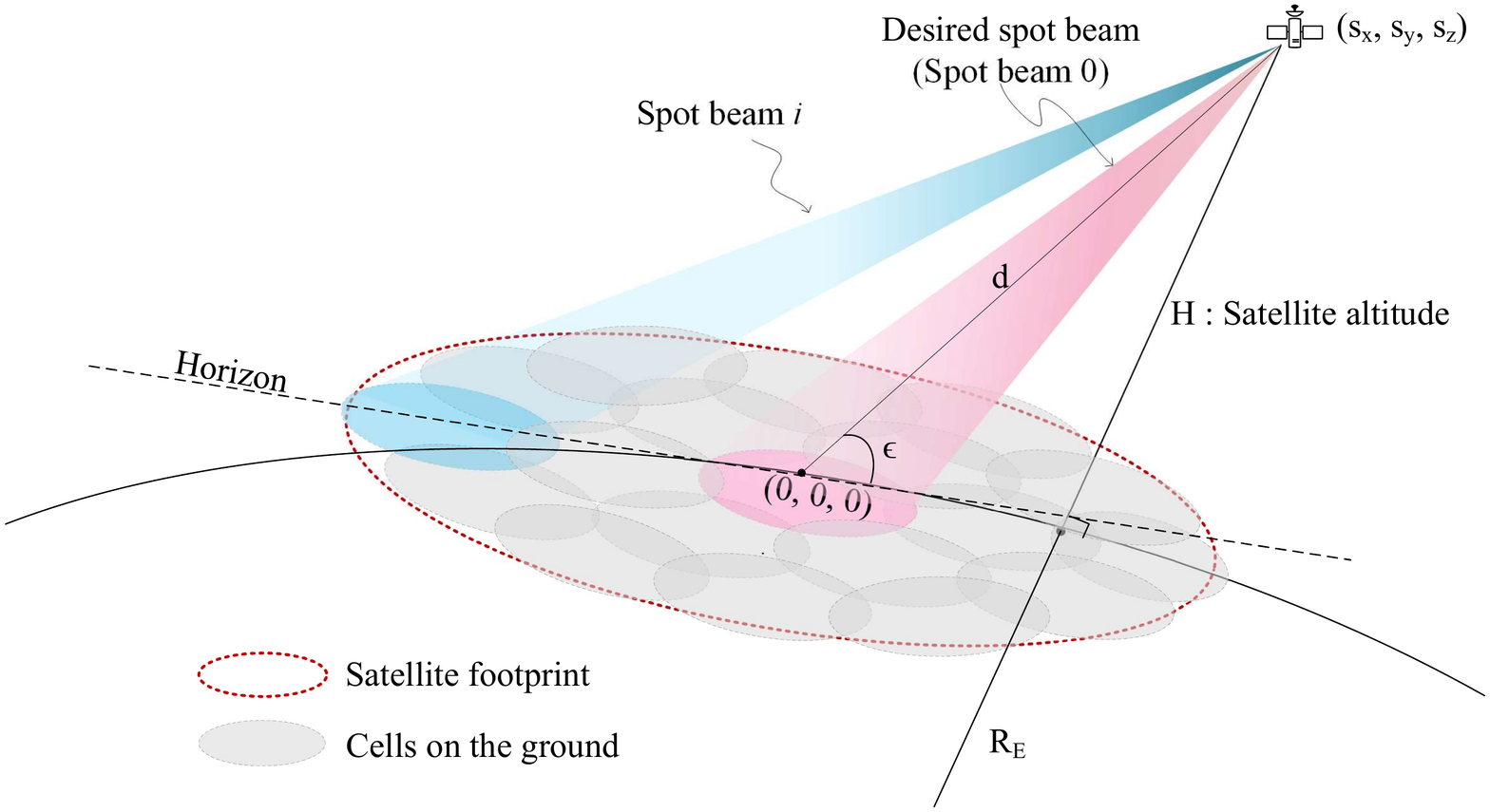} 
		\caption{A LEO satellite located at $(s_x, s_y, s_z)$ with altitude $H$ and elevation $\epsilon$ delivers downlink to ground users with multiple onboard transmitters, each of which steers a spot beam to illuminate a cell on the ground, collectively comprising the satellite footprint. In addition to receiving desired downlink signals, ground users also incur interference from neighboring spot beams.}
		\label{fig:footprint}
\end{figure*}

We consider a single \leo satellite serving downlink to several ground users simultaneously through the use of multiple spot beams.
Each spot beam is formed by a dedicated transmitter onboard the satellite, which provides service to users within its \textit{cell} on the ground.
As illustrated in \figref{fig:footprint}, cells are tessellated to form the total \textit{footprint} of the satellite. 
For our formulation, we assume all spot beams operate over the same frequency spectrum (i.e., full frequency reuse), and for brevity, do not incorporate ground user receive beam gain, since it acts on desired signal and interference, but this can be included straightforwardly.

At a given instant, suppose the satellite is located at a position $(s_x, s_y, s_z)$ relative to some origin on the surface of the Earth, which can be written as
\begin{align}
(s_x, s_y, s_z) = (d \cos \epsilon \cos \Phi ,d \cos \epsilon \sin \Phi, d \sin \epsilon),
\end{align}
where $\epsilon$ and $\Phi$ are the elevation and azimuth angles of the satellite, respectively, and $d$ is the absolute distance (or \textit{slant distance}) to the satellite.
The slant distance $d$ can be expressed in terms of the satellite altitude $H$ and its elevation angle $\epsilon$ as
\begin{align}
d = \sqrt{\radiusearth^2 \sin^2 \epsilon + H^2 + 2 H \radiusearth} - \radiusearth \sin \epsilon \label{eq:distance},
\end{align}
where $\radiusearth \approx 6378$ km is the radius of the Earth. 

Let $\numbeams$ be the number of spot beams maintained by the satellite, where each spot beam is driven by a dedicated transmitter onboard the satellite with total conducted transmit power $\powertx$.
We denote $G_i\parens{\phi,\theta}$ as the gain of the $i$-th spot beam toward some azimuth $\phi$ and elevation $\theta$ relative to its steering direction, where $i = 0, \dots, \numbeams-1$.
One can consider the case where each spot beam is steered toward the center of the cell it serves, as illustrated in \figref{fig:footprint}.
Note that this formulation can be used for both dish and array-based satellite antennas. 
Let $x_i$ be the transmitted symbol from the $i$-th onboard transmitter, where $\ev{\bars{x_i}^2} = 1$.
Transmissions by each spot beam will inflict interference onto ground users served by the other $\numbeams - 1$ beams, since practical beam patterns naturally leak energy in undesired directions.
Given the overwhelming distance between the satellite and a ground user relative to the separation between onboard antennas, a desired signal and the corresponding $\numbeams - 1$ interference signals experience approximately the same propagation channel $h$ and same path loss $\Ppl$.
As such, we can write the received symbol of a user being served by the $0$-th spot beam as
\begin{align}
y_0 
=& \underbrace{\sqrt{\Ptx \cdot \Ppl^{-1} \cdot G_0(\phi_0, \theta_0)} \cdot h \cdot x_0}_{\mathsf{desired~signal}} \ + \ \underbrace{\sum_{i=1}^{\numbeams-1} \sqrt{\Ptx \cdot \Ppl^{-1} \cdot G_i(\phi_i, \theta_i)} \cdot h \cdot x_i}_{\mathsf{interference}} \ + \ n_0, \label{eq:signals}
\end{align}
where $h$ is the propagation channel and $n_0 \sim \distcgauss{0}{\noisevar}$ is additive noise.
Here, $\parens{\phi_i,\theta_i}$ is the relative azimuth-elevation of the ground user relative to the steering direction of the $i$-th spot beam. 
Consequently, the degree of interference incurred by a ground user depends on its location and the steering directions of the $\numbeams$ spot beams (i.e., the cell placement).

We model sky-to-ground propagation with the \sr channel model \cite{newsimple}, where the channel magnitude is an \sr random variable distributed as 
\begin{align}
\bars{h} \sim \distSR{b,m,\Omega},
\end{align}
whose \gls{pdf} is defined as
\begin{align}
f_{|h|}(x; b,m,\Omega) 
= \frac{x}{b} \left(\frac{2b m}{2bm + \Omega}\right)^{m}  \exp\left(-\frac{x^2}{2b}\right) {}_1\mathcal{F}_1 \left(m, 1, \frac{\Omega x^2 }{2b (2bm + \Omega)}\right),
\label{eq:pdfsr} 
\end{align}
where ${}_1\mathcal{F}_1(\cdot, \cdot, \cdot)$ is the confluent hypergeometric function \cite{tablesofintegrals}, namely
\begin{align}
{}_1\mathcal{F}_1(a, b, x) = \sum_{i=0}^{\infty} \frac{(a)_i}{i!(b)_i}x^i,
\end{align}
with $(a)_i = a(a+1) \dots  (a+i-1)$ is the Pochhammer symbol \cite{mathhandbook}.
Based on actual measurements \cite{loo,channel_measure}, the \sr channel model accurately captures both \los and \nlos propagation in a Rician fashion and incorporates random fluctuations of each, caused by obstructions such as buildings, trees, and vegetation \cite{newsimple}.  
The three parameters of the \sr channel model can be summarized as:
\begin{itemize}
    \item $\Omega$ being the average power of the \los component;
    \item $2b$ being the average power of the \nlos component;
    \item $m$ being the fading order dictating the general shape of distribution.
\end{itemize}
With this presented downlink system model, we derive and characterize key performance metrics in the next section.

\section{Characterizing Performance in Shadowed Rician Channels}\label{sec:sr-channel}
Using the system model presented in the previous section, we aim to characterize key performance metrics of the system, most notably \gsnr and \gsinr, along with \gsir and \ginr, which can drive system design, as we will highlight herein.
In doing so, we establish several relations between linearly-related \sr random variables, allowing us to describe these performance metrics in terms of the \sr channel parameters $(b,m,\Omega)$.

\subsection{Desired Signal Power and SNR}
With the magnitude of the channel modeled as an \sr random variable $\bars{h} \sim \distSR{b,m,\Omega}$, the channel power gain follows a squared \sr (\ssr) distribution as
\begin{align}
\bars{h}^2 \sim \distSSR{b,m,\Omega},
\end{align}
with its \gls{pdf} given as \cite{newsimple}
\begin{align}
f_{|h|^2}(y; b, m, \Omega) = \frac{1}{2b} \parens{\frac{2bm}{2bm + \Omega}}^{m} \exp\parens{-\frac{ y}{2b}}{}_1\mathcal{F}_1 \parens{m, 1, \frac{\Omega y}{2b (2b m+\Omega)}}. \label{eq:ssrpdf} 
\end{align}
Its \cdf is quite involved but can be expressed as \cite{closed_sr}
\begin{align}
	F_{|h|^2}(Y; b, m, \Omega) = \frac{1}{2b} \parens{\frac{2bm}{2bm + \Omega}}^{m}
	x \Phi_2 \parens{1-m, m; 2; -\frac{x}{2b_0}, -\frac{m x}{2 b m + \Omega}},
	 \label{eq:ssrcdf} 
\end{align}
where $\Phi_2$ is the bivariate confluent hypergeometric function defined as \cite{mathhandbook,bi_variate}
\begin{align}
	\Phi_2 (a, a'; c; w, z) = \sum_{k=0}^\infty \frac{(a)_k}{k! (c)_k} w^k {}_1\mathcal{F}_1 (a', c+k, z).
\end{align}

From \eqref{eq:signals}, we can write the power of the desired signal received by a ground user served by the $0$-th spot beam as
\begin{align}
\powerdes = \powertx \cdot \Ppl^{-1} \cdot G_0(\phi_0, \theta_0) \cdot \bars{h}^2,
\end{align}
which itself is a random variable linearly related to $\bars{h}^2$, since all other terms are deterministic for a given ground user location.
To describe $\powerdes$, we derive the following theorem to establish the relationship between \sr and \ssr random variables. 

\begin{theorem} \label{thm:scaled-ssr}
    If $X \sim \distSR{b,m,\Omega}$ and $Y = k \cdot X^2$ for $k >0$, then 
    \begin{align}
    Y \sim \distSSR{k \cdot b,m,k \cdot \Omega}.
    \end{align}
    \begin{proof}
        The \pdf of $Y$ is obtained by plugging in $x = \sqrt{\frac{y}{k}}$ into $f_Y(y) = f_X(x) \frac{dx}{dy}$, leading to
        \begin{align}
        f_Y(y)
        &= f_{|h| = X} \parens{x=\sqrt{\frac{y}{k}}; b,  m, \Omega} \frac{dx}{dy} \label{eq:thm_1-pre} \\
        &= \frac{1}{2kb} \parens{\frac{2kbm}{2kbm + k\Omega}}^{m} \exp\parens{-\frac{ y}{2kb}}{}_1\mathcal{F}_1 \parens{m, 1, \frac{k\Omega y}{2kb (2kb m+k\Omega)}}   \\
        &= f_{|h|^2=Y}(y; kb, m, k\Omega) ,
        \label{eq:thm_1}
        \end{align}
        where $f_{|h|}(\cdot)$ and $f_{|h|^2}(\cdot)$ are given in \eqref{eq:pdfsr} and \eqref{eq:ssrpdf}, respectively. 
    \end{proof}
\end{theorem}

\begin{corollary}
    Two \ssr random variables $Y_1 \sim \distSSR{b_1,m_1,\Omega_1}$ and $Y_2 \sim \distSSR{b_2,m_2,\Omega_2}$ are linearly related as $Y_1 = k \cdot Y_2$ for $k > 0$ if
    \begin{gather}
    m_1 = m_2 \quad \textrm{and} \quad \frac{b_1}{b_2} = \frac{\Omega_1}{\Omega_2} = k.
    \end{gather}
    \label{co:linear-ssr}
\end{corollary}

With $\bars{h} \sim \distSR{b,m,\Omega}$ distributed as an \sr random variable, \thmref{thm:scaled-ssr} states that $\powerdes$ is an \ssr random variable that can be inferred directly as
\begin{align}
\powerdes \sim \distSSR{\dot{b},\dot{m},\dot{\Omega}},
\end{align}
with shadowing parameters scaled accordingly as
\begin{align}
\dot{b} &= \Ptx \cdot \Ppl^{-1} \cdot G_0(\phi_0, \theta_0) \cdot b, \\ 
\dot{m} &= m, \\ 
\dot{\Omega} &= \Ptx \cdot \Ppl^{-1} \cdot G_0(\phi_0, \theta_0)  \cdot \Omega.
\end{align}
Notice that, when scaling the \ssr random variable, the fading order $m$ remains unchanged; only the average powers of the \los and \nlos components have changed.

Perhaps more meaningful than desired signal power in dictating system performance is \gsnr, which is also a random variable and can be written as
\begin{align}
\msnr = \frac{\powerdes}{\noisevar} = \overline{\msnr} \cdot \bars{h}^2,
\end{align}
where we use $\lsnr$ to denote the large-scale \gsnr without random channel variations as
\begin{align} \label{eq:msnr}
\overline{\msnr}= \frac{\Ptx \cdot \Ppl^{-1} \cdot G_0(\phi_0, \theta_0)}{\noisevar}. 
\end{align}
Since $\msnr$ is linearly related to $\bars{h}^2$, \thmref{thm:scaled-ssr} states that \snr follows an \ssr distribution tied to that of the channel $h$ as 
\begin{align}
\msnr \sim \distSSR{\lsnr \cdot b, m, \lsnr \cdot \Omega}.
\end{align}
In this setting, given the presence of multi-beam interference, \gsnr only partly dictates system performance.
Nonetheless, it is important to realize that the distribution of $\msnr$ sets the upper bound on system performance.
As such, under \sr channels, it is critical that the system be designed so that $\lsnr$ is sufficiently high with a high probability for any user needing service.
In the next subsection, we characterize interference.

\subsection{Interference Power and INR}
The power of interference incurred by a user served by the $0$-th spot beam can be written as
\begin{align}
\powerint 
&= \sum_{i=1}^{\numbeams-1} \powertx \cdot \Ppl^{-1} \cdot G_i(\phi_i, \theta_i) \cdot \bars{h}^2,
\end{align}
which depends on the channel gain $\bars{h}^2$ and the gain of each interfering spot beam in the direction of the user.
As with desired signal power, \thmref{thm:scaled-ssr} states that interference power is an \ssr random variable distributed as
\begin{align}
\powerint \sim \distSSR{\bar{b},\bar{m},\bar{\Omega}},
\end{align}
with shadowing parameters scaled as
\begin{align}
\bar{b} &= \sum_{i=1}^{\numbeams-1} \powertx \cdot \Ppl^{-1} \cdot G_i(\phi_i, \theta_i) \cdot b, \\ 
\bar{m} &= m, \\ 
\bar{\Omega} &=  \sum_{i=1}^{\numbeams-1} \powertx \cdot \Ppl^{-1} \cdot G_i(\phi_i, \theta_i) \cdot \Omega.
\end{align}
\ginr is an important quantity for communication systems plagued by interference since it indicates if the system is noise-limited ($\minr \ll 0$ dB) or interference-limited ($\minr \gg 0$ dB).
Like $\msnr$, $\minr$ is linearly related to $\bars{h}^2$ as
\begin{align}
\minr = \frac{\powerint}{\noisevar} = \linr \cdot \bars{h}^2,
\end{align}
where $\linr$ is the large-scale \ginr capturing the leakage of each interfering beam onto the ground user being served,
\begin{align}
\linr = \frac{\powertx \cdot \Ppl^{-1} \cdot \sum_{i=1}^{\numbeams-1} G_i(\phi_i, \theta_i)}{\noisevar}.
\end{align}
\thmref{thm:scaled-ssr} straightforwardly describes $\minr$ as an \ssr random variable distributed as
\begin{align}
\minr \sim \distSSR{\linr \cdot b, m, \linr \cdot \Omega}.
\end{align}
Notice that $\linr$ is solely a function of system parameters: transmit power, path loss, noise power, and the sum spot beam gain toward the ground user.
This naturally introduces the challenge of cell planning and spot beam steering to mitigate the effects of interference (deliver a low $\linr$) without degrading coverage.
Although $\minr$ is a random variable, engineers can use $\linr$ to ensure a given ground user sees below some level of interference with certain probability, based on the distribution of the \sr channel.
As mentioned in the introduction, the beam gain observed by users on the ground is a function of the elevation angle of the satellite since its pattern appears distorted as it projects onto the surface of the Earth.
The satellite elevation angle adds to the complexity of cell planning and spot beam steering, which we investigate further in \secref{sec:simulation-results}.

\subsection{SIR and SINR} \label{subsec:sir-sinr}
The strength of a desired signal and that of interference are both useful metrics on their own, but combining the two provides truer indications of system performance.
To begin, we consider \gsir, that is the ratio of desired signal power to that of interference, which can in fact be written deterministically as
\begin{align}
\msir 
= \frac{\powerdes}{\powerint} 
= \frac{\msnr}{\minr} 
= \frac{\lsnr}{\linr} 
= \frac{G_0(\phi_0, \theta_0)}{\sum_{i=1}^{\numbeams-1} G_i(\phi_i, \theta_i)}. 
\label{eq:sir}
\end{align}
While desired signal power $\powerdes$ and interference power $\powerint$ are both \ssr random variables, it is important to note that they are fully correlated, both depending on the same random variable $\bars{h}^2$.
Recall, this is due to the fact that the stochastics seen by the signal of the $0$-th spot beam are also seen by the signals of the other spot beams considering they propagate along the same path from the satellite to a given ground user. 
From \eqref{eq:sir}, it is clear that \gsir only depends on the position of the ground user and the steering directions of the $\numbeams$ spot beams.

Unlike \gsir, the \gsinr of the system is indeed a random variable defined as
\begin{align}
\msinr 
= \frac{\powerdes}{\noisevar + \powerint} 
= \frac{\msnr}{1 + \minr} 
= \frac{\lsnr}{\frac{1}{\bars{h}^2} + \linr} 
\leq \min\parens{\msnr,\msir},
\end{align}
which does not follow the \ssr distribution and cannot be easily described.
However, by considering that $\msinr$ is upper-bounded by the minimum of $\msnr$ and $\msir$, useful results emerge. 
In noise-limited regimes (i.e., when $\linr$ is low), \gsinr can be approximated by \gsnr, meaning it is approximately distributed as
\begin{align}
\msinr \overset{\linr \to 0}{\sim} \distSSR{\lsnr \cdot b, m, \lsnr \cdot \Omega}. 
\end{align}
On the other hand, when interference-limited (i.e., when $\linr$ is high), \gsinr is approximated by \gsir, from which it follows that
\begin{align}
\msinr \overset{\linr \to \infty}{=} \msir = \frac{G_0(\phi_0, \theta_0)}{\sum_{i=1}^{\numbeams-1} G_i(\phi_i, \theta_i)}. 
\label{eq:sinr-to-sir}
\end{align}
Notice that, while the true level of interference $\minr$ is a random variable, engineers can rely on $\linr$---which is based system parameters---to gauge conditions where $\msinr$ can be approximated by $\msnr$ or $\msir$ with certain probability.
Additionally, since $\msinr$ is upper-bounded by these two quantities, engineers can potentially leverage the fact that $\msir$ is deterministic for cell planning and to design beam steering solutions that ensure that the system design does not bottleneck performance, regardless of shadowing conditions of the channel $h$.
With key performance metrics characterized in this section, we evaluate their stochastics in the next section to facilitate statistical analyses of \leo satellite systems.

\section{A Useful Approximation on Fading Order} \label{sec:integer-m}
In the previous section, we characterized downlink \gsnr, \ginr, \gsir, and \gsinr of a multi-beam \leo system under the \sr channel model.
As mentioned in the introduction and made evident in the previous two sections, the statistics of \sr and \ssr random variables generally involve complex expressions and special functions, and their moments (e.g., expectation) cannot be stated concisely.
This complicates statistical analysis of these key performance metrics.
In this section, we show that statistically characterizing \sr and \ssr random variables simplifies when the fading order $m$ of an \sr random variable is an integer.

\subsection{Probability of Outage}
The probability that a desired signal's quality falls below some threshold---or the probability of outage---is an important quantity for evaluating and characterizing a communication system.
In this interference setting, $\msinr$ falling below some threshold is presumably the key metric of interest.
However, as mentioned, the distribution of $\msinr$ is not straightforward, and even the \cdf of $\msnr$ itself is quite involved, as shown in \eqref{eq:ssrcdf}.
To express the \pdf and \cdf of $\msnr$ in a closed-form without the use of infinite power series, we rely on the following theorem and corollaries.

\begin{theorem} \label{thm:m-integer}
    When the fading order $m$ is an integer, the \gls{pdf} of an \ssr random variable $Y \sim \distSSR{b,m,\Omega}$ as shown in \eqref{eq:ssrpdf} can be simplified as
    \begin{align}
    \tilde{f}_Y(y;b,m,\Omega) =&\frac{1}{2b} \left(\frac{2b m}{2b m + \Omega}\right)^{m}   \exp\parens{-\frac{my}{ 2b m + \Omega }}  
    \sum_{i=0}^{m -1} \frac{(m - 1)! }{(m-1-i)!(i!)^2}\parens{\frac{\Omega y}{2b (2b m + \Omega) }}^i,
    \label{eq:integer-pdf} 
    \end{align}
    and its \gls{cdf} as shown in \eqref{eq:ssrcdf} can be simplified as
    \begin{align}
    \tilde{F}_Y(y) =&\left(\frac{2b m}{2b m + \Omega}\right)^{m-1}  
    \sum_{i=0}^{m -1} \frac{(m - 1)! }{(m-1-i)!(i!)^2}\parens{\frac{\Omega }{2bm}}^i  \parens{i! - \gamma\parens{i+1, \frac{m y }{2 b m + \Omega}}}, \label{eq:integer-cdf}
    \end{align}
    where $\gamma(a,x) = \int_{0}^{x} e^{-t}t^{a-1} \ dt$ is the unnormalized incomplete Gamma function \cite{mathhandbook}.
    
    \begin{proof}
        Rewriting the confluent hypergeometric function for integer $m \geq 1$ as a polynomial via Kummer's transform \cite{mathhandbook}, we have 
        \begin{align}
        {}_1\mathcal{F}_1(m, 1,\Omega) 
        &= e^{\Omega} {}_1\mathcal{F}_1(1-m, m, -\Omega) 
        =e^{\Omega}\sum_{i=0}^{m-1} \frac{(m-1)! \cdot \Omega^i}{(m-1-i)! \cdot (i!)^2}.
        \label{eq:confluent_sim}
        \end{align}
        which, along with algebra, yields \eqref{eq:integer-pdf}. 
        Similarly, using \eqref{eq:confluent_sim}, the \gls{cdf} of $Y$ is obtained as 
        \begin{align}
        \tilde{F}_Y(y) 
        &= \frac{1}{2b} \left(\frac{2b m}{2b m + \Omega}\right)^{m} \int_0^y \exp\parens{-\frac{m}{ 2b m + \Omega }t} 
        \sum_{i=0}^{m -1} \frac{(m - 1)! }{(m-1-i)!(i!)^2}\parens{\frac{\Omega t }{2b (2b m + \Omega) }}^i dt \label{eq:integer-cdf-begin}  \\
        &=\frac{1}{2b} \left(\frac{2b m}{2b m + \Omega}\right)^{m}  
        \sum_{i=0}^{m -1} \frac{(m - 1)! }{(m-1-i)!(i!)^2}\parens{\frac{\Omega }{2b (2b m + \Omega) }}^i 
        \int_0^y \exp\parens{-\frac{m}{ 2b m + \Omega }t}  t^i dt   \\
        &=\left(\frac{2b m}{2b m + \Omega}\right)^{m-1}  
        \sum_{i=0}^{m -1} \frac{(m - 1)! }{(m-1-i)!(i!)^2}\parens{\frac{\Omega }{2bm}}^i  \parens{i! - \gamma\parens{i+1, \frac{m y }{2 b m + \Omega}}}.  
        \end{align}
    \end{proof}
\end{theorem}

With \thmref{thm:m-integer}, we can represent the \gls{pdf} and \gls{cdf} of an \ssr random variable under integer fading order $m$ without an infinite power series in either expression. 
Using this, the probability of \gsnr outage is directly computed as
\begin{align}
\mathbb{P}\parens{\msnr \leq \gamma} = \tilde{F}_Y\parens{\gamma \cdot {\overline{\msnr}}^{-1}} 
\label{eq:outage}.
\end{align}
Although this \gsnr outage is an underestimate on the probability of \gsinr outage, it provides a closed-form expression for quantifying outage probability that offers convenience both numerically and analytically.
This probability of \gsnr outage is especially useful in settings where interference is low, such as under less aggressive frequency reuse.
Additionally, since $\msinr \leq \min\parens{\msnr,\msir}$ and the fact that $\msir$ is deterministic for a given design, engineers can calculate the probability that the design is not noise-limited by computing
\begin{align}
\prob{\msnr \leq \msir} = \tilde{F}_Y\parens{\msir \cdot \lsnr\inv}.
\label{eq:cdf_sir}
\end{align}
While this equality only holds when $m$ is an integer, later in this section we show that rounding $m$ to an integer often has minor impacts on the distribution, meaning it can be reliably used to closely approximate the \pdf and \cdf of \ssr random variables, even when $m$ is not an integer.

\begin{figure}[t]
    \centering
    \includegraphics[width=\linewidth,height=0.4\textheight,keepaspectratio]{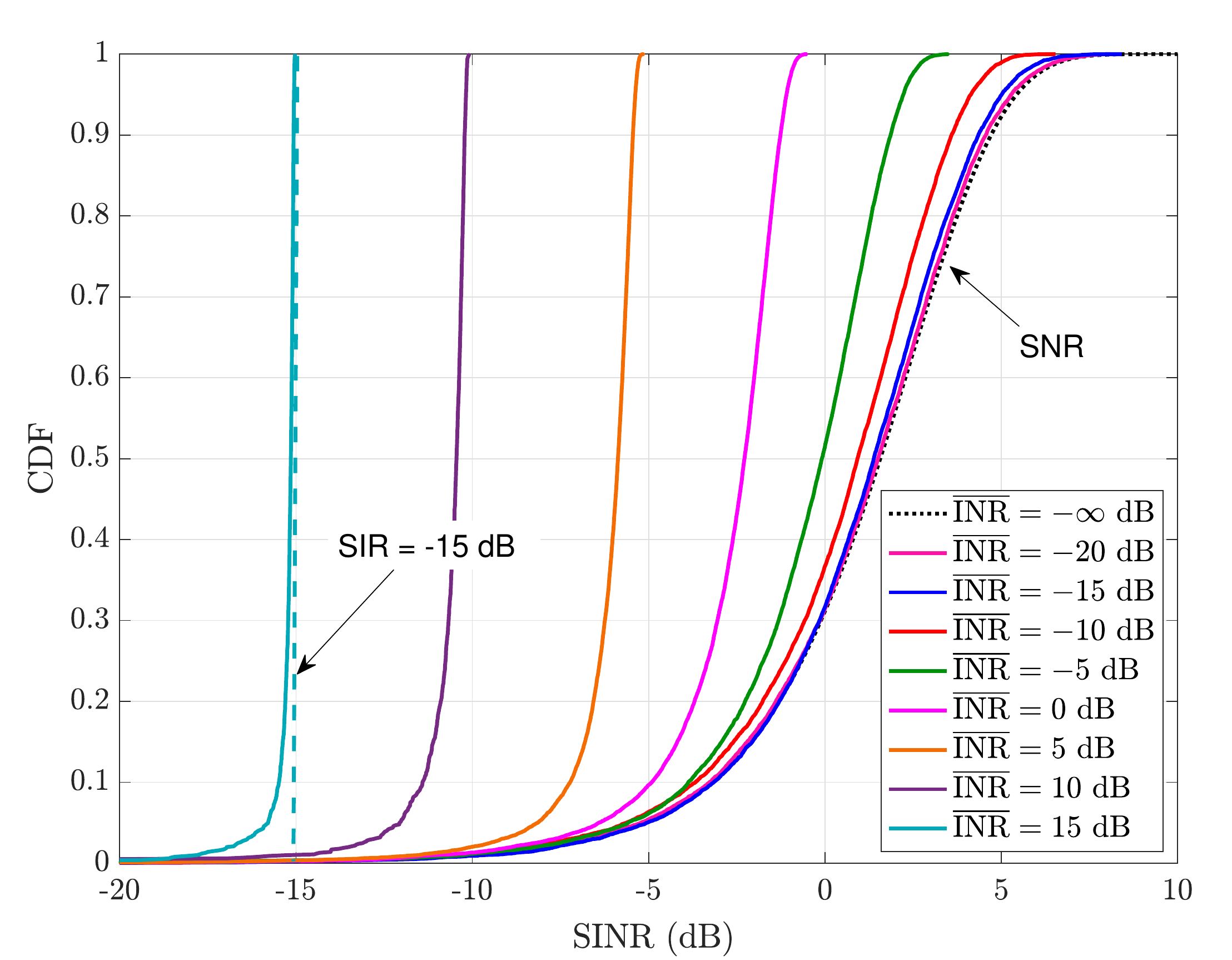}
    \caption{The empirical \cdf of $\msinr$ for various $\linr$ where $\lsnr = 0$ dB and under light shadowing conditions (with integer $m$). The dotted black line where $\linr = -\infty$ dB corresponds to the numerical \cdf of $\msnr$ using \eqref{eq:integer-cdf}, to which the distribution of $\msinr$ converges at low $\linr$. At high $\linr$, $\msinr$ converges to $\msir$ as evident by the increasing steepness of its \cdf.} 
    \label{fig:cdf}
\end{figure}

To illustrate how the \cdf of $\msnr$ in \eqref{eq:integer-cdf} can be used to approximate that of $\msinr$, consider \figref{fig:cdf}.
For various $\linr$, we draw realizations of $\bars{h}^2 \sim \distSSR{b,m,\Omega}$ under light shadowing (which we will describe in detail shortly \cite{newsimple,loo}) and calculate the resulting $\msinr$ as
\begin{align}
\msinr = \frac{\lsnr \cdot \bars{h}^2}{1 + \linr \cdot \bars{h}^2},
\end{align}
where we fix $\lsnr = 0$ dB.
We plot the empirical \cdf of $\msinr$ and compare it against the \cdf of $\msnr = \lsnr \cdot \bars{h}^2$ based on \eqref{eq:integer-cdf}.
When $\linr$ is sufficiently low (e.g., $\linr \leq -15$ dB), the distribution of $\msnr$ reliably approximates that of $\msinr$. 
Therefore, if a satellite system can estimate $\lsnr$ and $\linr$, which are based solely on system parameters, and has an estimate of the \sr channel statistics, it can obtain an approximate distribution of $\msinr$, assuming $\linr$ is sufficiently low.
As remarked earlier, if $\linr$ is sufficiently high, $\msir$ is a good approximation of $\msinr$, in which case it is deterministic based on beam steering and cell placement, as described by \eqref{eq:sir}.
This can be observed in \figref{fig:cdf} as the \cdf of \sinr trends toward $\msir = -15$ dB as $\linr = 15$ dB (recall, $\lsnr = 0$ dB in this example).

\subsection{Expected SNR and INR}
In addition to probability of outage, it is also useful to examine the mean \gsnr and \ginr of a system. 
Recall, the mean of an \ssr random variable is highly involved for general $m$ \cite{newsimple}; the following corollary can be used to express it in an intuitive closed-form when the fading order $m$ is an integer.

\begin{corollary} \label{cor:mean-ssr}
    The mean of $Y \sim \distSSR{b,m,\Omega}$ when $m$ is an integer is
    \begin{align}
    \ev{Y} = 2 \cdot b + \Omega. 
    \label{eq:corol-mean}
    \end{align}
    \begin{proof}
        The expected value of $Y$ is derived as 
        \begin{align}
        \E[Y] &= \int_{0}^{\infty}  y \tilde{f_Y}(y) dy \label{eq:integer-m1}\\ 
        &= \frac{1 }{2b} \left(\frac{2b m}{2b m + \Omega}\right)^{m}  
        \sum_{i=0}^{m -1} \frac{(m - 1)! }{(m-1-i)!(i!)^2}\parens{\frac{\Omega }{2b (2b m + \Omega) }}^i \int_0^\infty \exp \parens{- \frac{m y}{2 b m + \Omega}}  y^{i+1} dy \label{eq:integer-m2}\\
        &\ea\frac{1}{2b} \left(\frac{2b m}{2b m + \Omega}\right)^{m} \parens{\frac{2 b m + \Omega}{m}}^2  \sum_{i=0}^{m -1} \frac{(m-1)!(i+1)}{(m-1-i)! i!} \parens{\frac{\Omega}{2 b m}}^i \\
        &\eb  (2 \cdot b + \Omega)
        \label{eq:integer-m3},
        \end{align}
        where $(\text{a})$ is obtained using \cite{tablesofintegrals}
        \begin{align}
        \int_{0}^{\infty} z^i e^{-\mu z} dz= i! \mu^{-i-1},
        \end{align}
        and $(\text{b})$ is derived using 
        \begin{align}
        \sum_{i=0}^{m -1} \frac{(m-1)!(i+1)}{(m-1-i)! i!} \parens{\frac{\Omega}{2 b m}}^i 
        &= \sum_{i=0}^{m -1} \binom{m-1}{i} (i+1)\parens{\frac{\Omega}{2bm}}^i \\
        &= \parens{1+ \frac{\Omega}{2bm}}^{m-2}\parens{1 + \frac{\Omega}{2b}}.
        \end{align}
    \end{proof}
\end{corollary}

\begin{corollary} \label{cor:m-1}
    In the special case when $Y \sim \distSSR{b,m,\Omega}$ with $m=1$, $Y$ follows the exponential distribution with \pdf and \cdf respectively as
    \begin{align}
    \tilde{f}_Y(y;b,1,\Omega) &=  \frac{1}{2b + \Omega} \cdot \exp\parens{-\frac{y}{ 2b  + \Omega }}  
    \label{eq:integer-pdf-1}, \\
    \tilde{F}_Y(y;b,1,\Omega) &= 1- \exp \parens{-\frac{y}{2b + \Omega}} 
    \label{eq:integer-cdf-1}.
    \end{align}
    The mean and variance of $Y$ are $\ev{Y} = 2 \cdot b + \Omega$ and $\var{Y} = \ev{Y}^2$, respectively.
\end{corollary}

Using \corref{cor:mean-ssr}, the expected \snr and \inr with integer fading order $m$ are simply
\begin{align}
\ev{\msnr} &= \lsnr \cdot \parens{2 \cdot b + \Omega} \label{eq:avg_snr}, \\
\ev{\minr} &= \linr \cdot \parens{2 \cdot b + \Omega} \label{eq:avg_inr}.
\end{align}
Unlike when $m$ is not an integer, these expected values are intuitively captured as the sum of the average powers of the \los and \nlos components of the \sr channel when $m$ is an integer.
For some $\lsnr$, $\linr$, and channel parameters $(b,\Omega)$, engineers can gauge the expected $\msnr$ and $\minr$ for any integer $m$.
Albeit limited, these quick calculations can be used by engineers to determine average performance of the system.
For instance, engineers can gauge if a particular user will be interference-limited on average or not, based solely on $\linr$---which depends only on system parameters---and an estimate of channel conditions $(b,\Omega)$.

\subsection{Approximating Fading Order as an Integer}
\begin{figure}[!t]
    \centering
    \includegraphics[width=\linewidth,height=0.4\textheight,keepaspectratio]{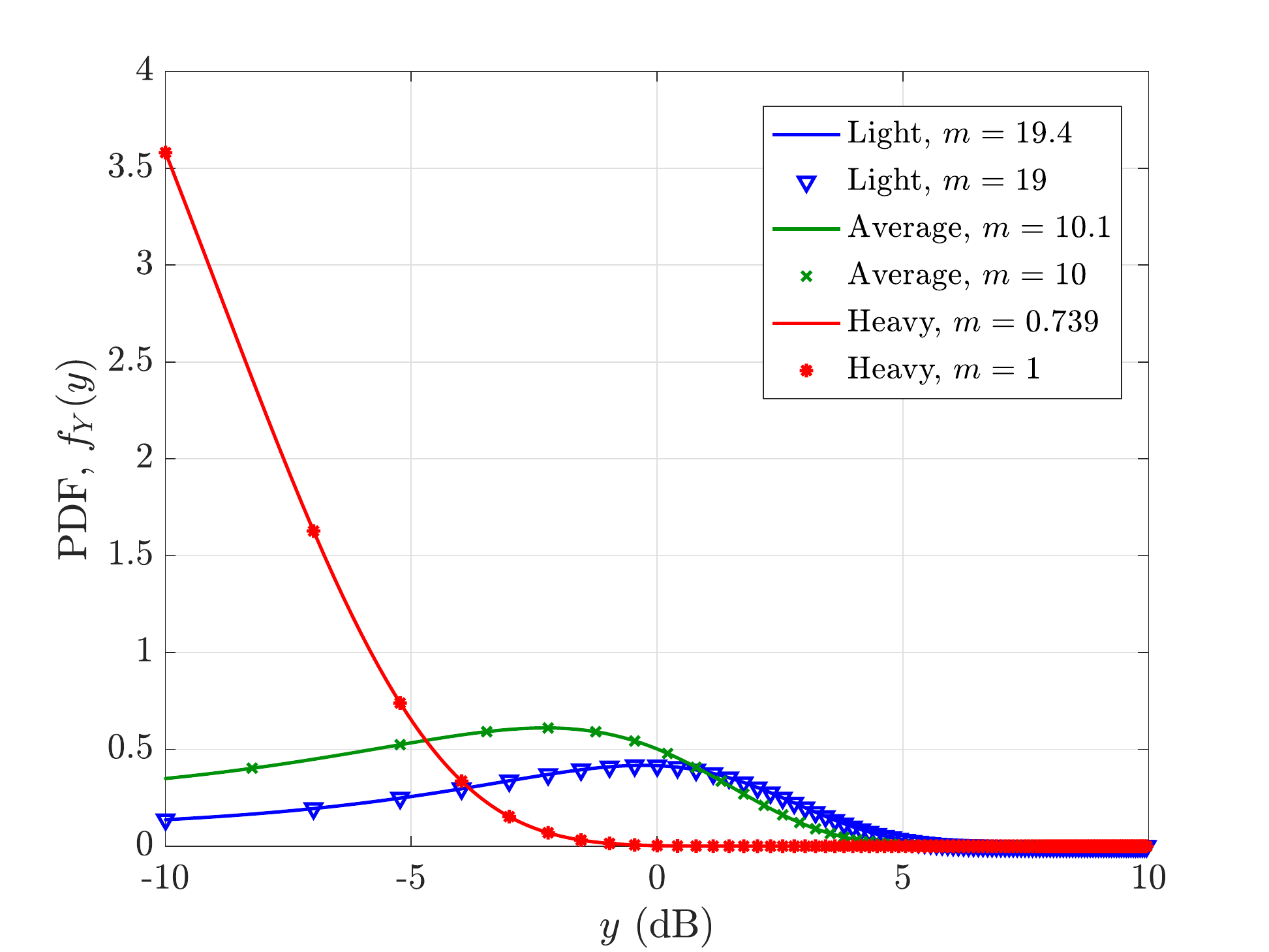} 
    \caption{The \gls{pdf} of an \ssr random variable $Y$ with integer and non-integer fading order $m$ for various levels of shadowing, as shown in \tabref{tab:sr-params} and fitted from measurements in \cite{newsimple,loo}.}
    \label{fig:integer-m}
\end{figure}

\thmref{thm:m-integer} and the consequent corollaries rely on the fading order $m$ being an integer.
In cases where $m$ is not an integer, approximating it as such can facilitate statistical analyses without deviating significantly from the original distribution with non-integer $m$.
In \figref{fig:integer-m}, we illustrate this with three different shadowing levels \cite{newsimple,loo}: light, average, and heavy, which are elaborated on and tabulated in \tabref{tab:sr-params} in the next section. 
The \pdf of the three shadowing levels with their true $m$ are shown as solid lines; markers indicate their counterparts with $m$ rounded to the nearest integer. 
Notice that $m$ varies from less than $1$ to over $19$, and all three distributions are extremely closely aligned---so much so that we had to use markers instead of separate lines to distinguish the two. 
With \glspl{pdf} extremely closely aligned with general $m$ and integer $m$, it is guaranteed that their statistics also be closely aligned. 
It is important to note that the parameters $\parens{b,m,\Omega}$ for these three shadowing levels were obtained by fitting the \sr distribution to channel measurements \cite{newsimple,loo}.
As such, one can reason that the effects of rounding $m$ to the nearest integer are even less pronounced in practice, since any statistical model fitted to measurements will inherently not perfectly align with reality.
Minute distributional differences invisible to the naked eye, therefore, are immaterial for most practical applications.
With all this being said, we believe \thmref{thm:m-integer} and \corref{cor:mean-ssr} can be used as fairly reliable and useful approximations for any \sr distribution by rounding fading order $m$ the nearest integer.

\begin{table}[t]
	\caption{System simulation parameters \cite{3gpp38811,3gpp38821}.}
	\centering
	\label{tab:sim-params}
	\begin{tabular}{| c| c| }
		\hline
		Altitude ($H$)	& $600$ km \\
		\hline
		Carrier frequency ($f_{\mathrm{c}}$) & $20$ GHz \\
		\hline
		System bandwidth	 & $400$ MHz  \\
		\hline
		Satellite transmit power & $4$ dBW/MHz \\
		\hline
		Maximum transmit beam gain & $38.5$ dBi \\
		\hline
		Ground user receiver type	 & VSAT \\
		\hline
		Maximum receive beam gain & $39.7$ dBi \\
		\hline
		Ground user noise figure & $1.2$ dB \\
		\hline 
		Ground user antenna temperature & $150$ K \\
		\hline
		Spotbeam cell boresight & Steered to the center of its cell \\
		\hline
		Cell radius & $10 $ km \\
		\hline
		Number of spot beams ($\numbeams$) & 19 \\
		\hline
	\end{tabular}
\end{table}

\begin{table}[t]
    \caption{\ssr parameters fitted from measurements \cite{newsimple,loo}.}
    \centering
    \label{tab:sr-params}
    \begin{tabular}{|c|c|c|c|}
        \hline
        Shadowing Level  & Light	&  Average  & Heavy \\
        \hline
        ${b}$ & $0.158$ & $0.126$ & $0.063$ \\
        \hline
        ${m}$ & $19.4$ & $10.1$ & $0.739$ \\
        \hline
        ${\Omega}$ & $1.29$ & $0.835$ & $8.97 \times 10^{-4}$ \\
        \hline
    \end{tabular}
\end{table}


\section{Performance Evaluation of a 20 GHz Multi-Beam LEO Satellite System\\in Shadowed Rician Channels} \label{sec:simulation-results}
In this section, we simulate a 20 GHz (Ka-band) multi-beam \leo satellite communication system and evaluate the impact various system parameters have on key performance metrics, namely \gsnr, \ginr, and \gsinr.
A summary of parameters used for simulation is listed in \tabref{tab:sim-params}, most of which are based on \cite{3gpp38811,3gpp38821} published by 3GPP.
We simulate a satellite at an altitude of $H = 600$ km equipped with $\numbeams = 19$ spot beams, each steered toward the center of its cell on the ground.
Cells are tessellated in a hexagonal fashion with a cell radius of $10$ km.
The gain delivered by the $i$-th spot beam to a user on the ground we model as a steerable dish antenna with gain pattern \cite{3gpp38811} 
\begin{align}
	G_i(\phi,\theta) = 
	\begin{dcases}
		1, & \zeta = 0^\circ \\
		4 \bars{\frac{J_1(ka \sin \zeta)}{ka \sin \zeta}}^2, & 0^\circ < |\zeta| \leq 90^\circ
	\end{dcases}
	\label{eq:beam}
\end{align}
where $\zeta = \arccos\parens{\cos \phi \cdot \cos \theta}$ is the absolute angle off antenna boresight, $J_1(\cdot)$ is the first-order Bessel function of the first kind, $a$ is the radius of the dish antenna, $k = 2 \pi / \lambda$ is the wave number, and $\lambda$ is the carrier wavelength. 
We model ground users as very small aperture terminals (VSAT) mounted on rooftops or vehicles with a maximum receive antenna gain of $39.7$ dBi, a noise figure of $1.2$ dB, and an antenna temperature of $150$ K (i.e., $G/T = 15.9$ dB/K) \cite{3gpp38821}. 
We assume ground users track their serving satellite to offer maximum receive gain and are associated to cells based on their locations.

\begin{figure}[t]
	\centering
	\includegraphics[width=\linewidth,height=0.4\textheight,keepaspectratio]{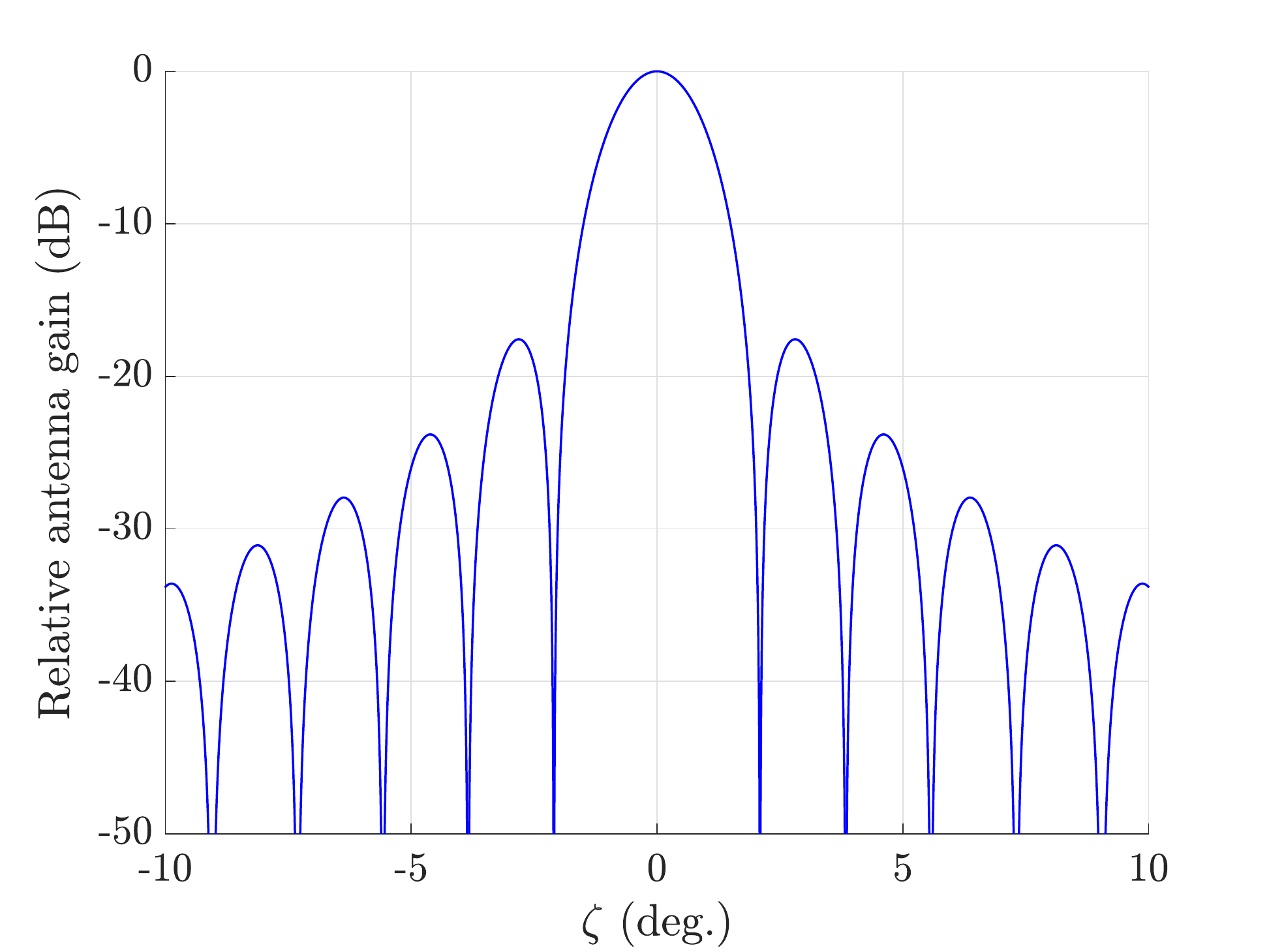}
	\caption{Normalized beam pattern of each high-gain dish antenna onboard the satellite as a function of angle off boresight $\zeta$.}
	\label{fig:dish-pattern}
\end{figure}

We consider \sr channels with three levels of shadowing---light, average, and heavy---whose parameters $(b,m,\Omega)$ are fitted from measurements in \cite{newsimple} and are shown in \tabref{tab:sr-params}.
Each onboard transmitter supplies $4$ dBW/MHz of power, and we simulate the system over a bandwidth of $400$ MHz. 
Path loss is modeled as the combination of free-space path loss and atmospheric attenuation as \cite{3gpp38811}
\begin{align}
	\mathrm{PL}(d,f_{\mathrm{c}}, \epsilon) = \mathrm{PL}_{\mathrm{FS}}(d, f_c) + \mathrm{PL}_{\mathrm{g}}(f_{\mathrm{c}}, \epsilon),
\end{align}
which is a function of propagation distance $d$, carrier frequency $f_{\mathrm{c}}$, and satellite elevation angle $\epsilon$.
Here, free-space path loss (in dB) is modeled as \cite{3gpp38811}
\begin{align}
	\todB{\mathrm{PL}_{\mathrm{FS}}(d, f_{\mathrm{c}})} = 32.45 + 20 \log_{10}(f_{\mathrm{c}}) + 20 \log_{10}(d),
\end{align}
which captures clutter loss and additional large-scale shadowing.
Absorption by atmospheric gases is modeled as \cite{itur_618_13,iturp676}
\begin{align}
	\mathrm{PL}_{\mathrm{g}} (f_{\mathrm{c}}, \epsilon) = \frac{A_{\mathrm{zen}}(f_{\mathrm{c}})}{\sin \epsilon},
\end{align}
where $A_{\mathrm{zen}}$ is a zenith attenuation given as $A_{\mathrm{zen}} = 0.9$ at a carrier frequency of $20$ GHz \cite{iturp676}.

\begin{figure*}[t]
    \centering
    \subfloat[Elevation of $\epsilon = 90^\circ$.]{\includegraphics[width=0.49\linewidth,height=\textheight,keepaspectratio]{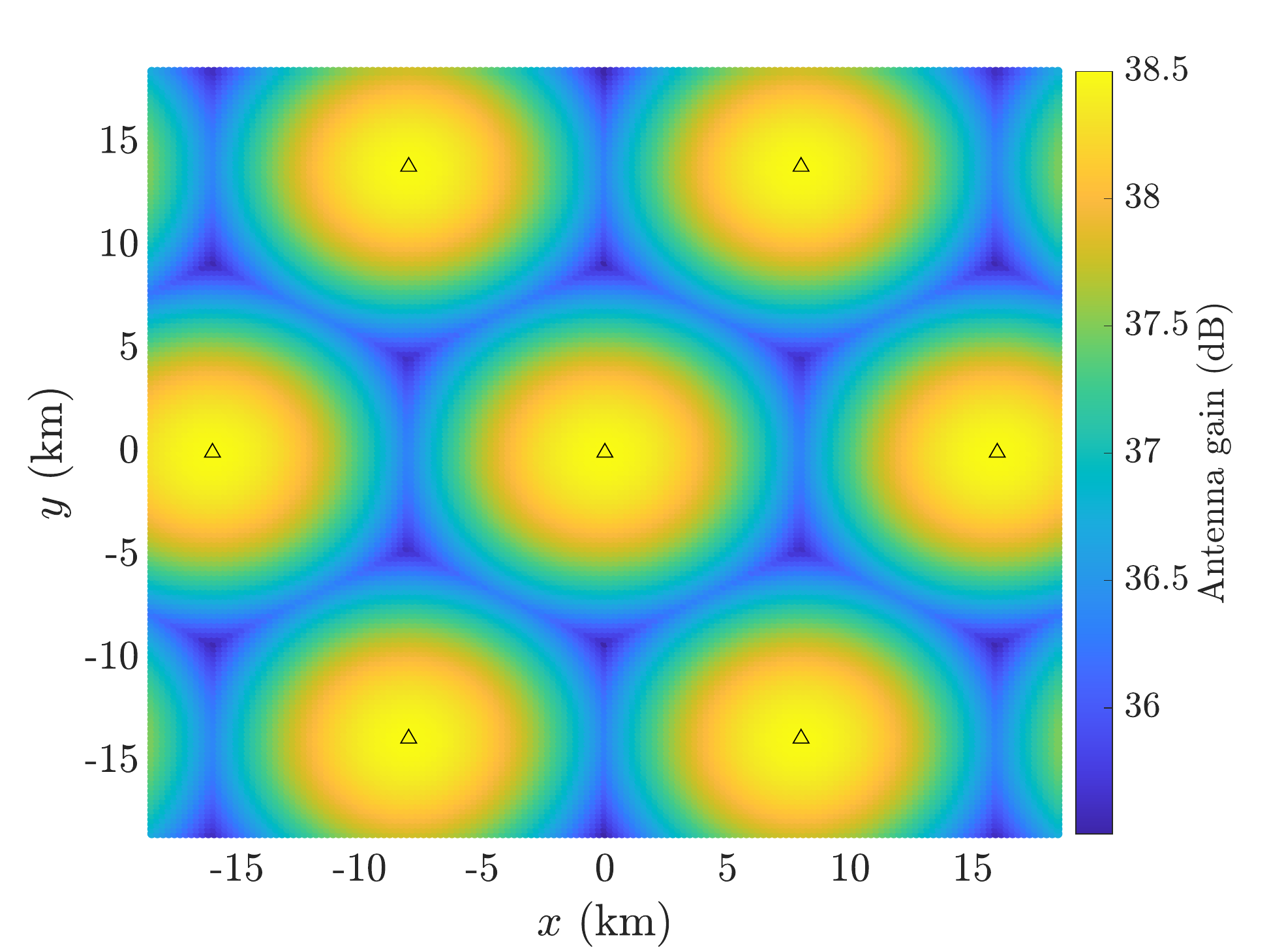}
        \label{fig:gain-90}}
    \hfill
    \subfloat[Elevation of $\epsilon = 45^\circ$.]{\includegraphics[width=0.49\linewidth,height=\textheight,keepaspectratio]{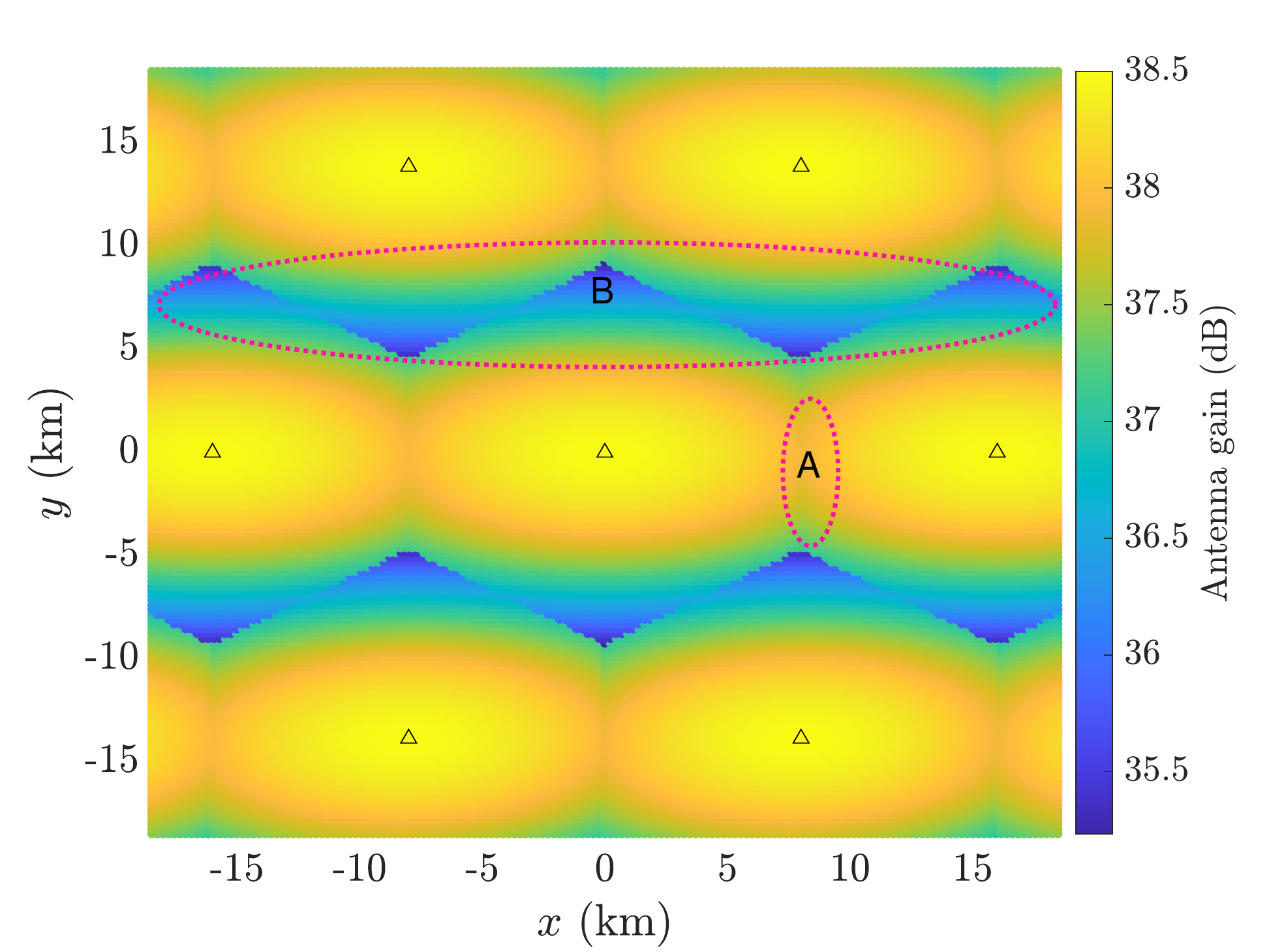}
        \label{fig:gain-45}}\\
    \caption{Delivered antenna gain as a function of ground user position for a satellite elevation angle of (a) $\epsilon = 90^\circ$ and (b) $\epsilon = 45^\circ$. Delivered beam gain distorts to a more elliptical shape at lower elevations, leading to less defined cell boundaries. Triangles denote cell centers. At an elevation of $\epsilon = 45^\circ$, ground users at the cell edge in region A enjoy $2$--$3$ dB higher beam gain than cell-edge users in region B, courtesy of the distorted beam shape.}
    \label{fig:gain-elevation}
\end{figure*}

\subsection{Effect of Elevation Angle on Antenna Gain}

We begin our system evaluation by highlighting the effect of satellite elevation angle on antenna gain delivered to ground users in \figref{fig:gain-elevation}.
In \figref{fig:gain-90}, the satellite is directly overhead the origin at an elevation of $\epsilon = 90^\circ$.
The satellite position and its elevation are relative to the center of the centermost cell, which is placed at the origin.
Plotting the observed antenna gain as a function of ground user location reveals the hexagonal arrangement of our cells; shown here are the six cells surrounding the centermost cell located at the origin.
The observed spot beam gain within each cell is nearly circular, which leads to well-defined cell boundaries.
Maximum gain is around $38.5$ dB with cell-edge users losing around $3$ dB of gain for this particular cell radius.
A user's distance from the center of its cell is a good indicator of the antenna gain it enjoys.
As elevation is decreased to $45^\circ$ (an increased presence in the $x$ direction), the satellite is closer to the horizon and the observed spot beam gain becomes more elliptical as it projects onto the surface of the Earth.
Consequently, the beam gain elongates along the $x$ dimension and tightens along the $y$ dimension. 
Cell boundaries are no longer as well-defined and user distance from the center of the cell is no longer a clear indicator of delivered beam gain. 
For instance, users in region A, enjoy near-maximal beam gain even though they are at the cell edge. 
Users in region B, also at the cell edge, see around $2$--$3$ dB less beam gain.
All these takeaways can be extrapolated for elevations between $45^\circ$ and $90^\circ$ and, through symmetry, beyond to $135^\circ$.


\begin{figure*}[!t]
    \centering
    \subfloat[Elevation of $\epsilon = 90^\circ$.]{\includegraphics[width=0.49\linewidth,height=\textheight,keepaspectratio]{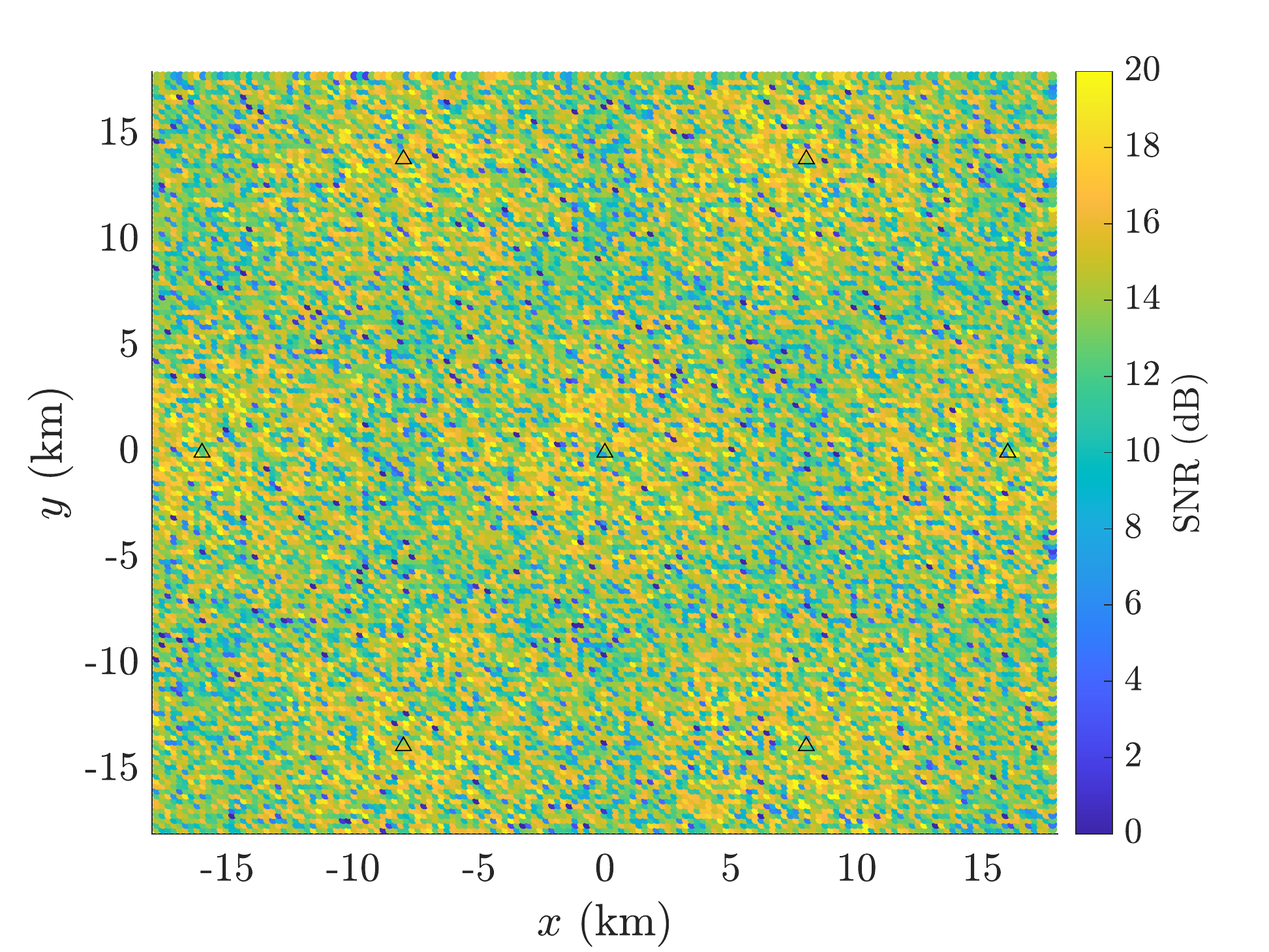}
        \label{fig:snr-90}}
    \hfill
    \subfloat[Elevation of $\epsilon = 45^\circ$.]{\includegraphics[width=0.49\linewidth,height=\textheight,keepaspectratio]{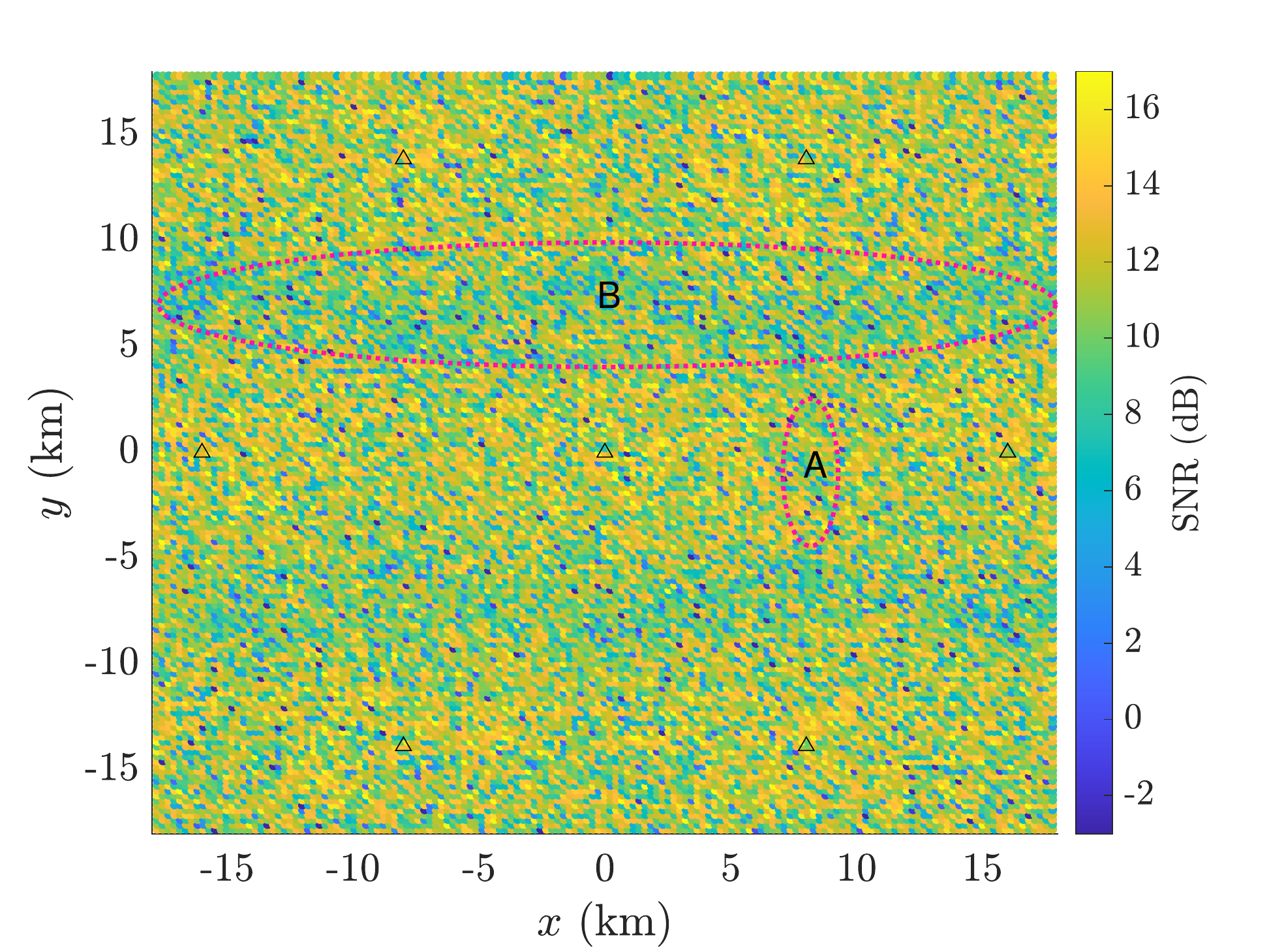}
        \label{fig:snr-45}}\\
    \caption{Received \gsnr as a function of user location under light shadowing for a satellite elevation angle of (a) $\epsilon = 90^\circ$ and (b) $\epsilon = 45^\circ$. User location is a good indicator for trends in \gsnr, but \sr channel stochastics can lead to deep fades even near the center of the cell. Triangles denote cell centers.}
    \label{fig:snr-elevation}
\end{figure*}

\subsection{SNR Distribution}

Now, instead of considering delivered beam gain, which is solely a function of cell placement and user location, we examine \gsnr as a function of user location in the presence of light shadowing (see \tabref{tab:sr-params}) in \figref{fig:snr-elevation}.
We again consider elevations of $\epsilon = 90^\circ$ and $\epsilon = 45^\circ$ in \figref{fig:snr-90} and \figref{fig:snr-45}, respectively.
The beam patterns observed before are apparent here at a high level as trends in \gsnr follow those seen in \figref{fig:gain-elevation}.
Users in region A tend to enjoy higher \gpsnr than those in region B.
In both cases, \gsnr tends to be higher near the center of each cell, with best-case users enjoying \gpsnr from around $15$ dB up to even $20$ dB.
Notice that \gpsnr observed at $\epsilon = 90^\circ$ are around $2$--$3$ dB higher than those at $\epsilon = 45^\circ$; this is attributed to increased slant distance $d$ (and hence path loss) at lower elevation angles.
Users enjoy \gsnr gains courtesy of constructive fading at the upper tail of the \sr channel distribution, particularly useful to those that observe lower $\lsnr$ at the cell edge, but more prominently, we see that shadowing can cause deep fades, regardless of user location.
Naturally, since users close to the center of the cell enjoy higher $\lsnr$, they are more robust to these deep fades but are not exempt from such.

\begin{figure}[!t]
    \centering
    \includegraphics[width=\linewidth,height=0.4\textheight,keepaspectratio]{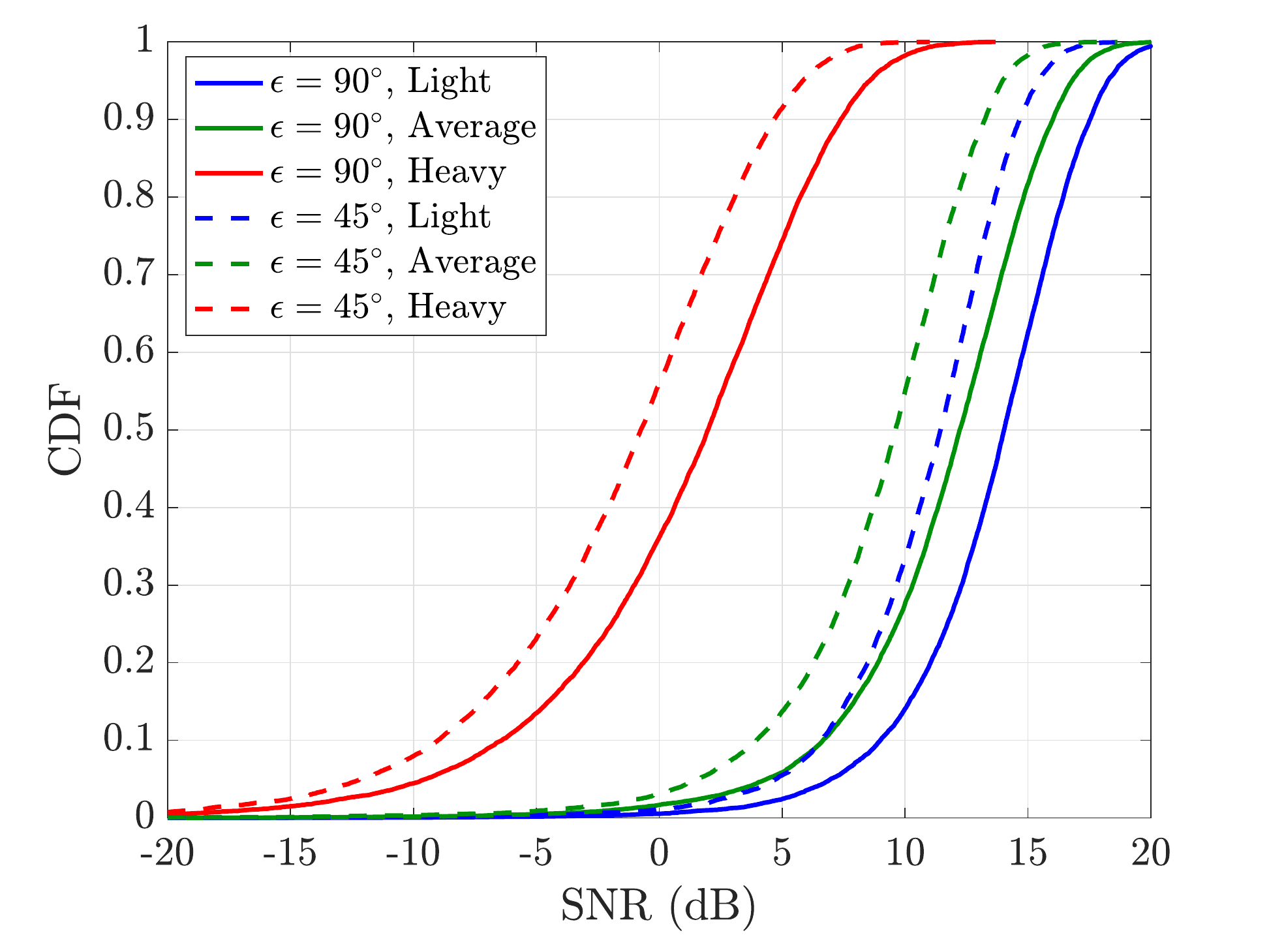}
    \caption{The \cdf of \gsnr for various shadowing levels and elevations of $\epsilon = 90^\circ$ and $\epsilon = 45^\circ$. As the satellite elevation angle falls from $90^\circ$ to $45^\circ$, users see a decrease of around $2$--$3$ dB in \gsnr distribution.} 
    \label{fig:snr-cdf}
\end{figure}

In \figref{fig:snr-cdf}, we plot the \cdf of \gsnr for various levels of shadowing and for elevation angles $\epsilon = 90^\circ$ and $\epsilon = 45^\circ$.
Here, the \cdf is taken across user locations within the center cell and over channel realizations.
Light shadowing conditions produce the highest \gsnr distribution, with median users enjoying around $\msnr = 14$ dB at $\epsilon = 90^\circ$ and just over $\msnr = 11$ dB at $\epsilon = 45^\circ$.
Worst-case users in light shadowing can suffer from deep fades that result in \gpsnr falling well below $5$ dB at both elevations.
As shadowing intensifies, the \gsnr distribution shifts leftward---a shift of about $12$ dB in the median from light to heavy shadowing---from which it is clear that shadowing level can severely impact performance.
Heavier shadowing produces a heavier lower tail and more variance overall. 
Since the effects of shadowing are independent of those due to elevation angle, there is a consistent $2$--$3$ dB gap in distribution between $\epsilon = 90^\circ$ and $\epsilon = 45^\circ$ across all three shadowing levels.
Again, interpolation between distributions can be used when considering elevations between $45^\circ$ to $135^\circ$.

\subsection{INR Distribution}

\begin{figure*}[!t]
    \centering
    \subfloat[Elevation of $\epsilon = 90^\circ$.]{\includegraphics[width=0.475\linewidth,height=\textheight,keepaspectratio]{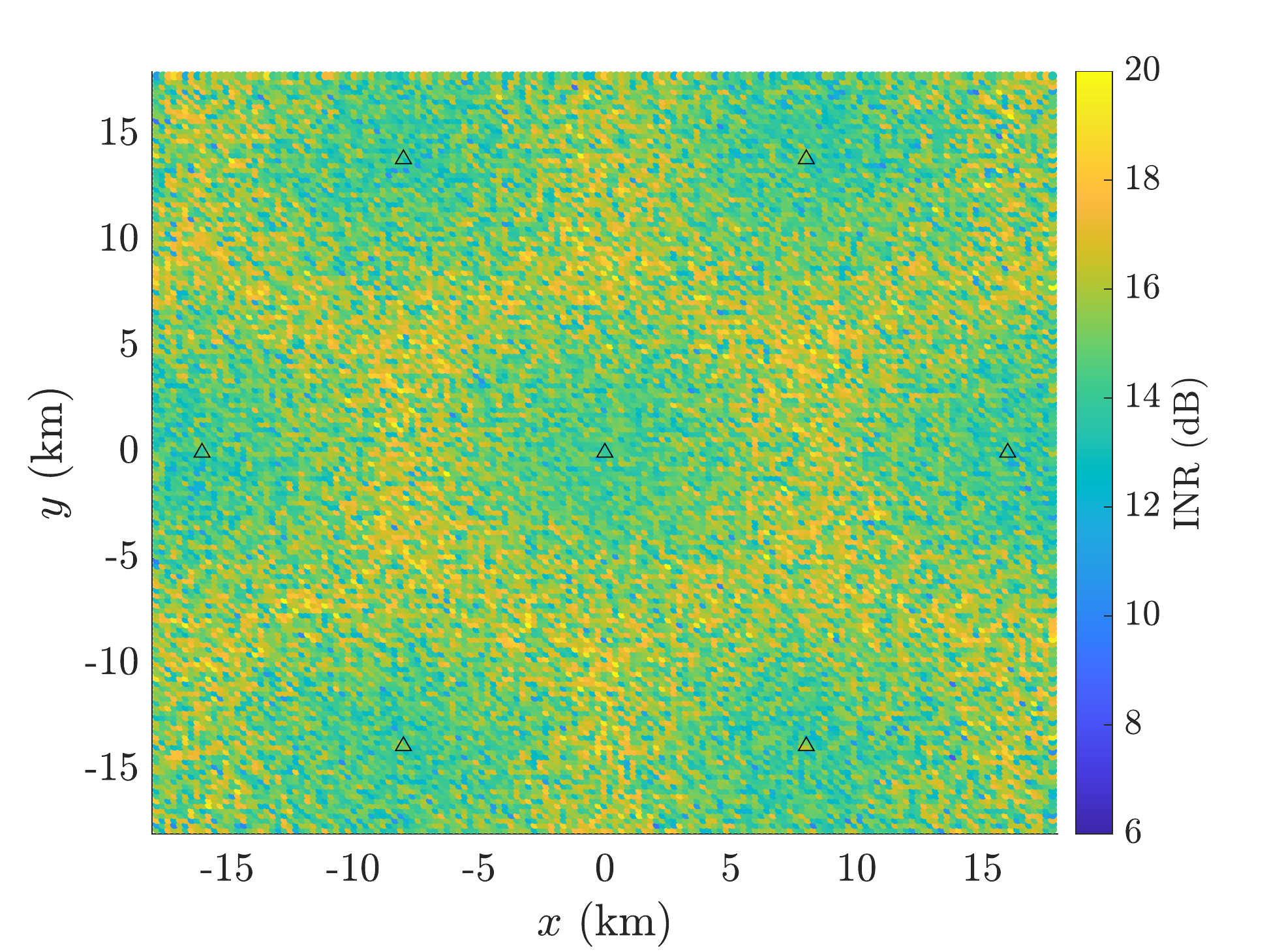} \hfill
        \label{fig:inr-90}}
    \subfloat[Elevation of $\epsilon = 45^\circ$.]{\includegraphics[width=0.475\linewidth,height=\textheight,keepaspectratio]{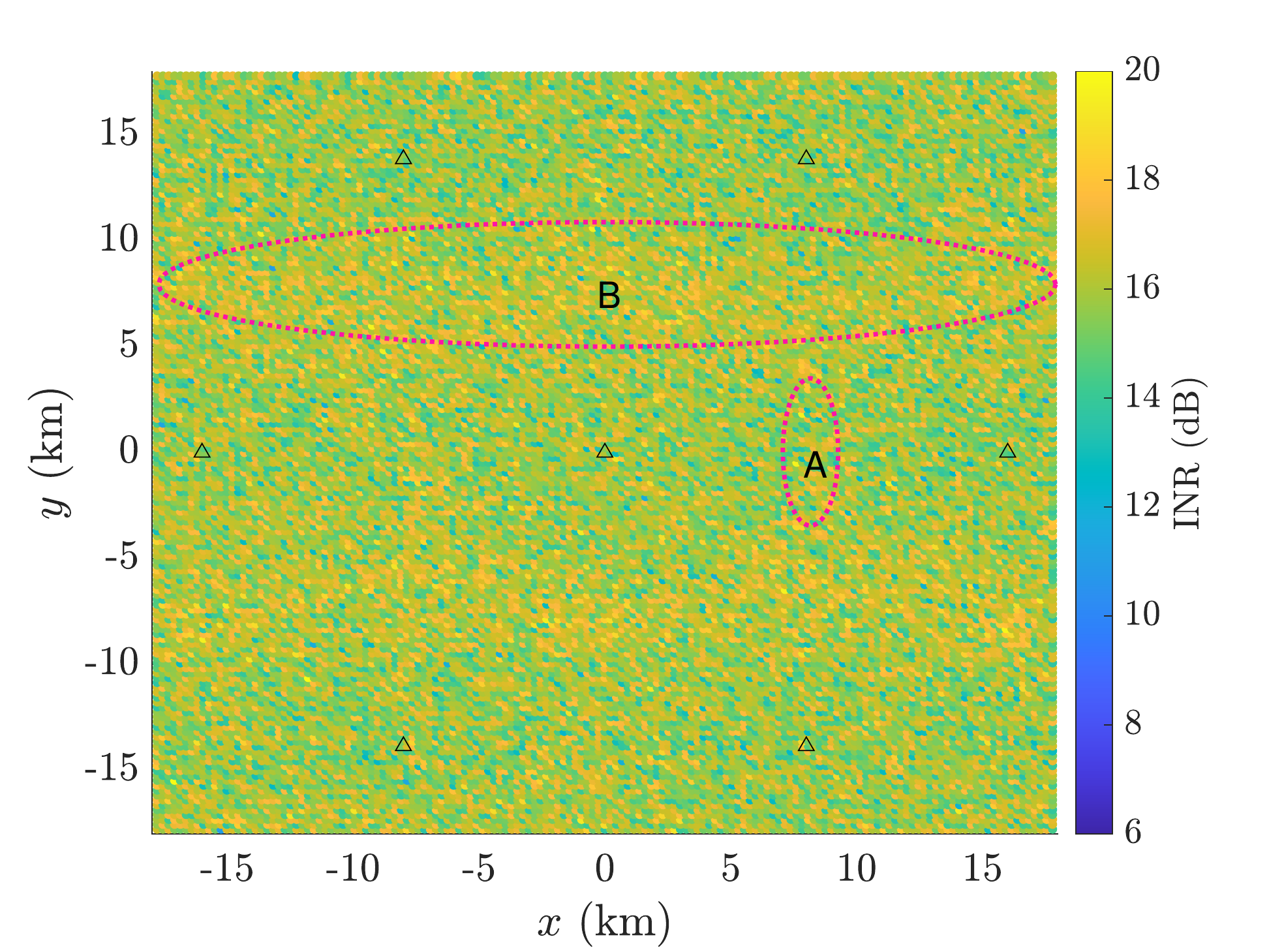}\hfill
        \label{fig:inr-45}}\\
    \caption{\ginr as a function of ground user location under light shadowing for a satellite elevation angle of (a) $\epsilon = 90^\circ$ and (b) $\epsilon = 45^\circ$. Users in region B tend to see more interference than those in region A at elevations below $90^\circ$. Triangles denote cell centers.}
    \label{fig:inr-el}
\end{figure*}

Having considered \gsnr, we now turn our attention to examining \ginr in a similar manner.
In \figref{fig:inr-el}, we plot a realization of \ginr as a function of ground user location for elevations of $90^\circ$ and $45^\circ$ under light shadowing.
At an elevation of $90^\circ$, \ginr typically ranges from around $10$ dB to upwards of $20$ dB.
Inverse to \gsnr, \ginr tends to increase as users approach the cell edge, where spot beam overlap is at its peak.
At an elevation of $45^\circ$, \ginr increases overall due to the distorted beam gain. 
Interestingly, we see that \ginr tends to be higher in region B compared to region A---opposite what was observed with \gsnr. 
This can be best explained by considering users located precisely at points A and B. 
A user at point A sees one dominant interferer (the spot beam serving the cell to the right of the center cell), whereas a user at point B sees the combination of two nearby interferers (the spot beams serving the two cells above the center cell).
Notice that the beam gains at these locations in \figref{fig:gain-45} differ by less than $3$ dB, meaning doubling the number of dominant interferers at point B will result in its total interference exceeding that at point A.

\begin{figure*}[!t]
	\centering
	\subfloat[Frequency reuse factor of one.]{\includegraphics[width=\linewidth,height=0.27\textheight,keepaspectratio]{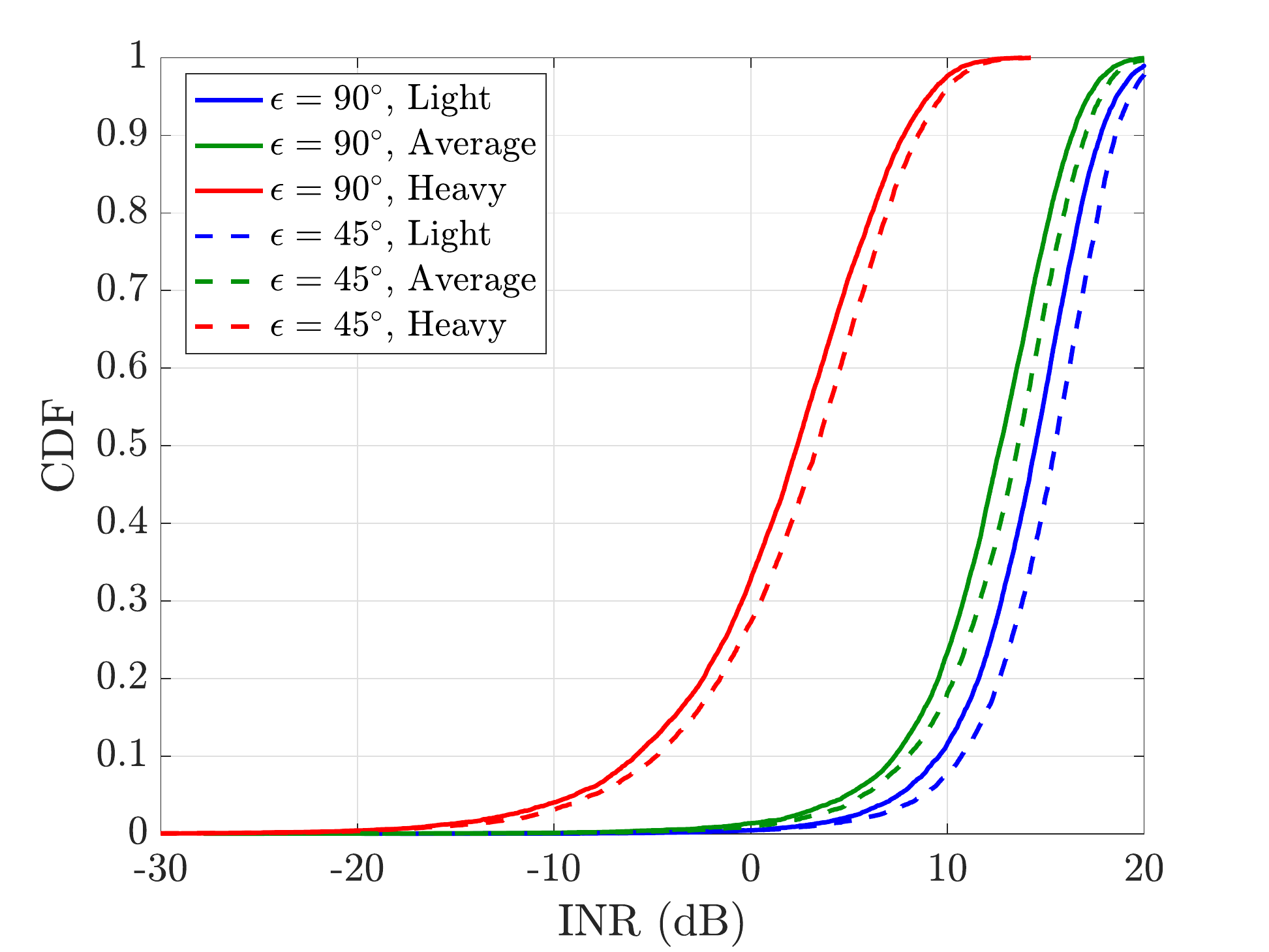}
		\label{fig:inr-cdf-a}}
	\subfloat[Frequency reuse factor of three.]{\includegraphics[width=\linewidth,height=0.27\textheight,keepaspectratio]{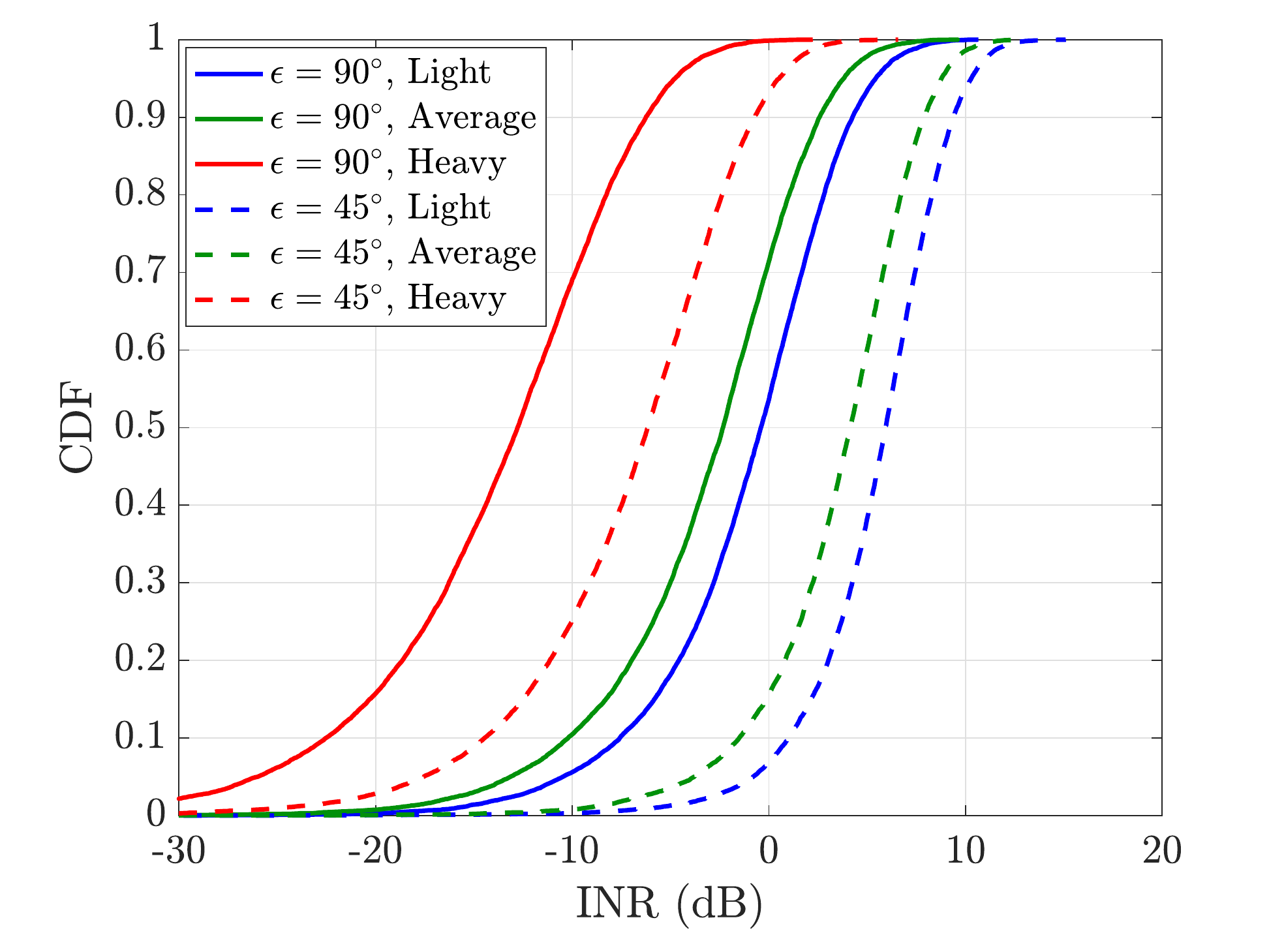}
		\label{fig:inr-cdf-b}}
	\caption{The \cdf of \ginr for various shadowing levels at elevations of $90^\circ$ and $45^\circ$ with (a) a frequency reuse factor of one and (b) a frequency reuse factor of three. As the frequency reuse factor increases, elevation angle plays a greater role in the degree of interference seen by users.}
	\label{fig:inr-cdf}
\end{figure*}

Thus far, we have assumed a frequency reuse factor of one.
Unlike \gsnr, \ginr is dictated by the particular frequency reuse factor since inter-beam interference is reduced as there is increased separation between beams operating on the same spectrum.
In \figref{fig:inr-cdf-a}, we plot the \cdf of \ginr for various levels of shadowing and for elevations $\epsilon = 90^\circ$ and $\epsilon = 45^\circ$, where the frequency reuse factor is one.
This is the \cdf of $\minr$ across ground user locations in the center cell in \figref{fig:inr-el}.
In \figref{fig:inr-cdf-b}, we plot that of \figref{fig:inr-cdf-a} except with a frequency reuse factor of three.
\figref{fig:inr-cdf-a} illustrates that increasing the shadowing level reduces interference between spot beams and shows that elevation plays a minor role in overall distribution---approximately a mere $1$ dB increase from $90^\circ$ to $45^\circ$.
With a frequency reuse factor of three, on the other hand, different conclusions are drawn.
As with a frequency reuse factor of one, overall interference reduces as shadowing intensifies with a frequency reuse factor of three.
Naturally, interference decreases as the frequency reuse factor is increased from one to three---here, by about $15$ dB.
Notice that, with a frequency reuse factor of one, the system tends to be interference-limited ($\minr > 0$ dB), except on occasion under heavy shadowing.
With a frequency reuse factor of three, however, the system is more often noise-limited.
When overhead at an elevation of $90^\circ$, even light shadowing has a median \ginr just less than $0$ dB.
As the elevation drops to $45^\circ$, the \ginr distribution shifts rightward by about $6$--$7$ dB, pushing the system to more often be interference-limited.
This shift in limitedness as the satellite traverses across the sky motivates the design of adaptable \leo satellite systems, which may sway from interference-limited to noise-limited and back to interference-limited within a minute or two.
In addition, these results emphasize that elevation, shadowing intensity, and frequency reuse factor should all be taken into account when evaluating the presence of spot beam interference and its distribution.

\subsection{SINR Distribution}

\begin{figure*}
    \centering
    \subfloat[Elevation of $\epsilon = 90^\circ$.]{\includegraphics[width=0.5\linewidth,height=0.29\textheight,keepaspectratio]{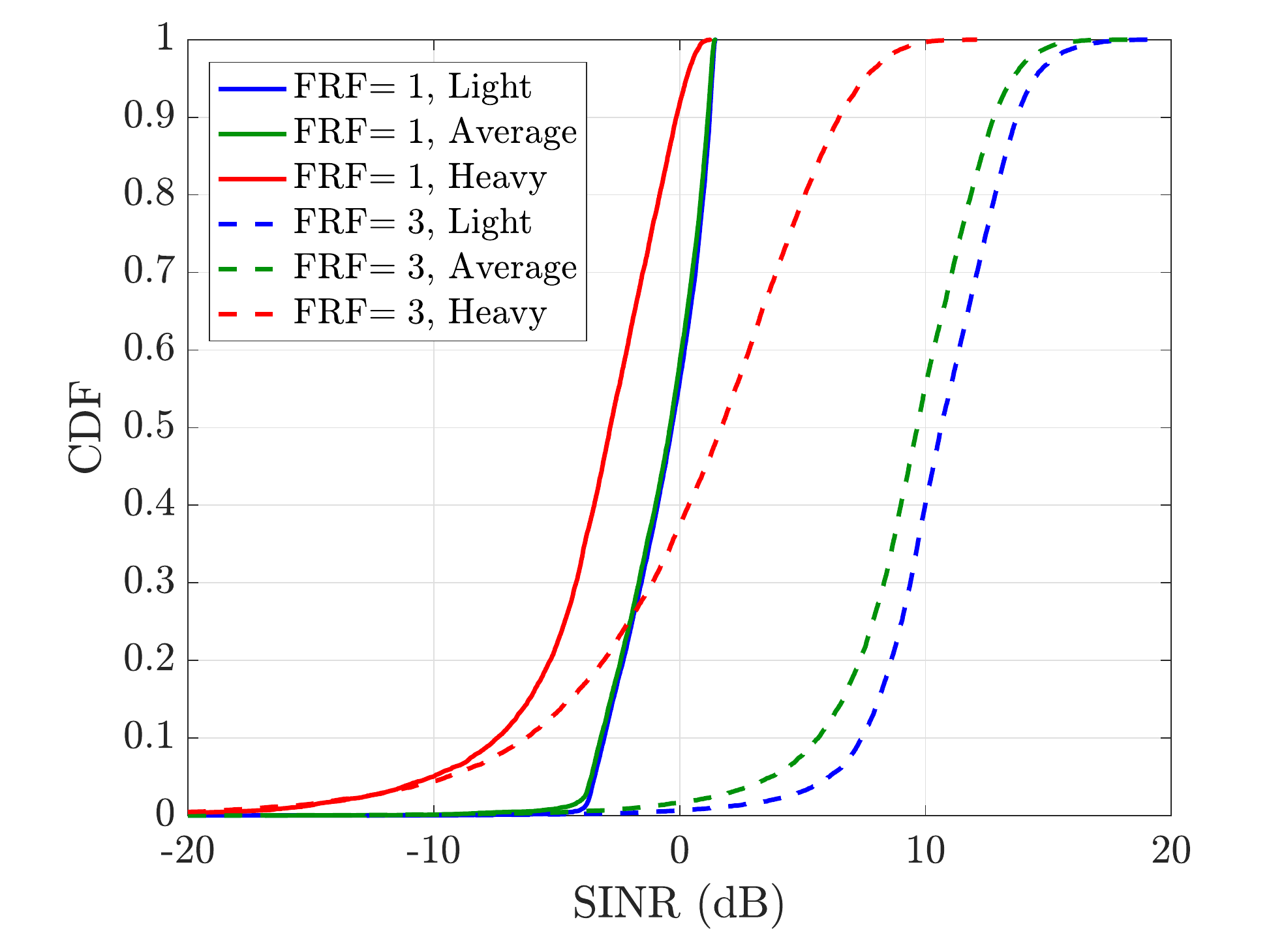}
        \label{fig:sinr-cdf-a}}
    \subfloat[Frequency reuse factor of three.]{\includegraphics[width=0.5\linewidth,height=0.29\textheight,keepaspectratio]{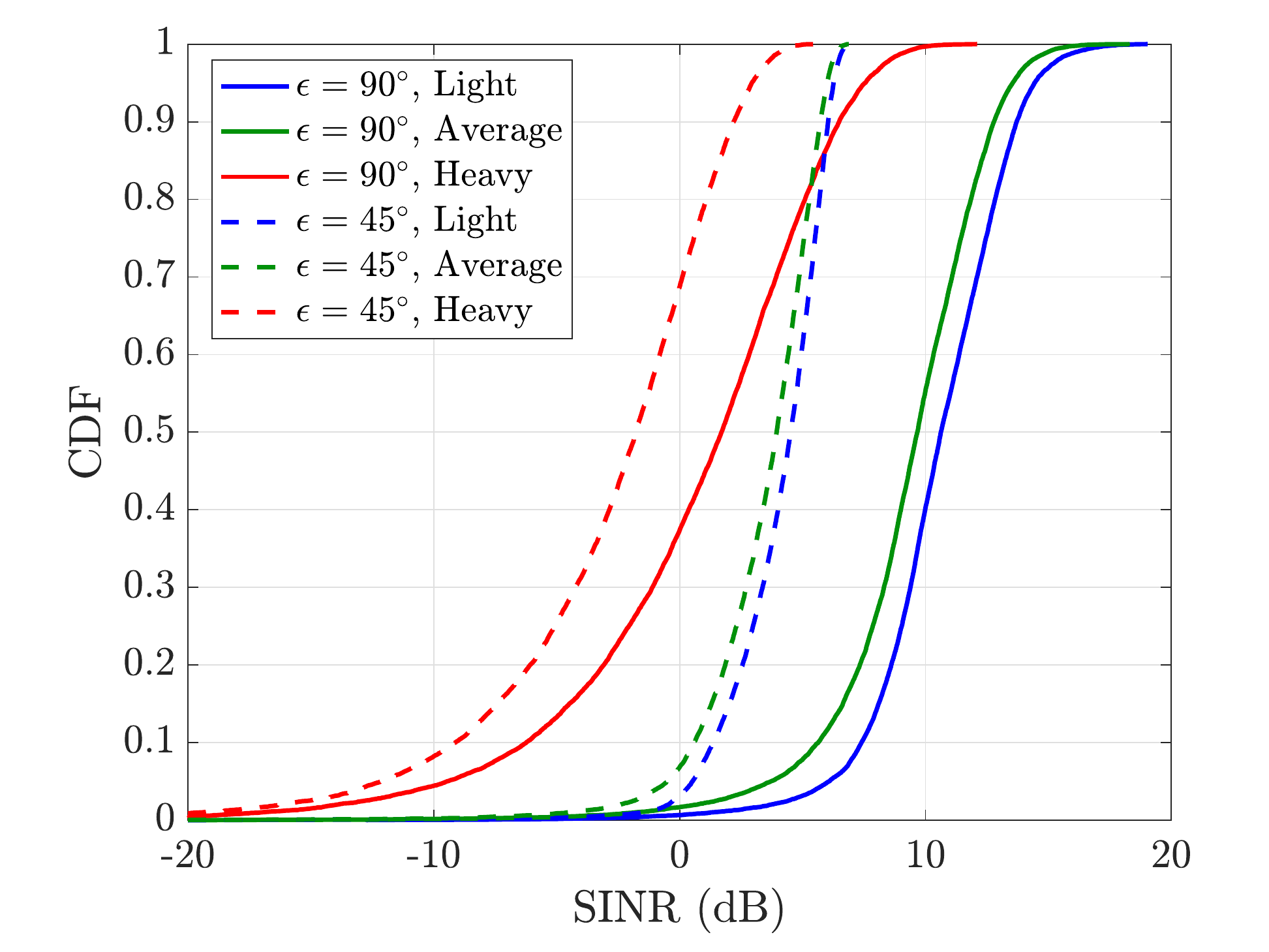}
        \label{fig:sinr-cdf-b}}
    \caption{(a) The \cdf of \gsinr under various shadowing levels for frequency reuse factors of one and three at an elevation of $\epsilon = 90^\circ$. (b) The \cdf of \gsinr under various shadowing levels for a frequency reuse factor of three at various elevation angles.}
    \label{fig:sinr-cdf}
\end{figure*}

To conclude our numerical evaluation, we now examine downlink \gsinr, the chief metric quantifying system performance.
In \figref{fig:sinr-cdf-a}, we similarly show the \cdf of \gsinr under various shadowing levels for frequency reuse factors of one and three at an elevation of $90^\circ$.
Under a frequency reuse factor of one, all three shadowing levels yield \gsinr distributions that lay largely below $0$ dB and with heavy lower tails.
System performance under heavy shadowing is particularly poor as over $90$\% of users experience $\msinr \leq 0$ dB.
The \gsinr distribution in light and average shadowing takes an interesting shape---a consequence of the system being primarily interference-limited, as noted before from \figref{fig:inr-cdf-a}.
Light and average shadowing yield nearly identical distributions.
This is due to the fact that both yield interference-limited conditions, meaning \gsinr can be approximated as \gsir, which is independent of the shadowing realization, as evidenced by \eqref{eq:sinr-to-sir}.
The sharp bend in these distributions can similarly be seen in \figref{fig:cdf} at high $\linr$.
Interference reduces as the frequency reuse factor is increased from one to three, improving median \gsinr by $5$ dB under heavy shadowing and by over $10$ dB under average or light shadowing.
This reduction in spot beam interference pushes the \gsinr distribution to levels that sustain communication and with less severe lower tails. 

In \figref{fig:sinr-cdf-b}, we fix the frequency reuse factor to three and highlight the effects of elevation angle.
As the satellite traverses from $90^\circ$ to $45^\circ$, the \gsinr distribution shifts leftward---a result of \gsnr decreasing and \ginr increasing as remarked before.
Under average and light shadowing, users see a reduction of around $6$ dB in median \gsinr at $45^\circ$ and, in heavy shadowing, experience $\msinr \leq 0$ dB $70$\% of the time.
As emphasized before, system performance can vary notably as the satellite traverses across the sky, largely due to the distorted beam shape observed by users on the ground.
This can lead to lower \gpsnr and higher interference, resulting in lower \gpsinr.
System performance improves as the satellite comes overhead and will degrade as it nears the horizon. 
With all of this happening over the course of a minute or two, appropriate measures should be taken to dynamically adapt the system based on satellite position and shadowing conditions.

\section{Conclusion and Future Directions} \label{sec:conclusion}

\leo satellite communication systems are evolving into a more prominent role connecting people and machines around the globe.
In this work, we analyzed multi-beam \leo satellite systems under the measurement-backed \sr channel model.
We derived key performance metrics including \gsnr, \ginr, \gsir, and \gsinr and provided a statistical characterization of each under \sr channels.
Our analyses and derivations can be useful tools for the statistical evaluation and the design of \leo satellite systems.
To facilitate this, we showed that rounding the \sr fading order to an integer can simplify expressions of \pdf, \cdf, and expectation, allowing researchers to more straightforwardly calculate probability of outage, for instance.
Then, we conducted a performance of evaluation of a 20 GHz multi-beam \leo system through simulation with practical system parameters and realistic antenna, channel, and path loss models.
Our results highlighted the effects of elevation angle, shadowing conditions, and frequency reuse factor on \gsnr, \ginr, and \gsinr, which motivates the need for frequency reuse factors above one and for systems that can adapt to varying conditions as the satellite traverses across the sky along its orbit.
Future work that can capitalize on the derivations and insights herein include optimal cell planning and spot beam design, along with other means to manage interference, potentially based on machine learning.
Naturally, strategic handover and scheduling will be paramount in successfully overcoming the constant orbiting of satellite base stations.
Finally, statistically characterizing entire networks of \leo satellites will be an essential stride toward validating the efficacy of these new wireless systems and the role they will play in next-generation connectivity.

\bibliographystyle{bibtex/IEEEtran}
\bibliography{refs}

\begin{thebibliography}{10}
\providecommand{\url}[1]{#1}
\csname url@samestyle\endcsname
\providecommand{\newblock}{\relax}
\providecommand{\bibinfo}[2]{#2}
\providecommand{\BIBentrySTDinterwordspacing}{\spaceskip=0pt\relax}
\providecommand{\BIBentryALTinterwordstretchfactor}{4}
\providecommand{\BIBentryALTinterwordspacing}{\spaceskip=\fontdimen2\font plus
\BIBentryALTinterwordstretchfactor\fontdimen3\font minus
  \fontdimen4\font\relax}
\providecommand{\BIBforeignlanguage}[2]{{%
\expandafter\ifx\csname l@#1\endcsname\relax
\typeout{** WARNING: IEEEtran.bst: No hyphenation pattern has been}%
\typeout{** loaded for the language `#1'. Using the pattern for}%
\typeout{** the default language instead.}%
\else
\language=\csname l@#1\endcsname
\fi
#2}}
\providecommand{\BIBdecl}{\relax}
\BIBdecl

\bibitem{globecom_mine}
E.~Kim, I.~P. Roberts, P.~A. Iannucci, and J.~G. Andrews, ``Downlink analysis
  of {LEO} multi-beam satellite communication in {Shadowed} {Rician}
  channels,'' in \emph{Proc. IEEE GLOBECOM}, Dec. 2021, pp. 01--06.

\bibitem{spaceX}
\BIBentryALTinterwordspacing
{SpaceX}. Non-geostationary satellite system attachment a technical information
  to supplement schedule {S}. Federal Communications Commission, Washington DC,
  USA, 2021. [Online]. Available:
  \url{https://fcc.report/IBFS/SAT-MOD-20181108-00083/1569860.pdf}
\BIBentrySTDinterwordspacing

\bibitem{kuiper}
\BIBentryALTinterwordspacing
{J. Hindin}. Technical {Appendix}, {Application} of {Kuiper} systems {LLC} for
  authority to launch and operate a non-geostationary satellite orbit system in
  {Ka-band} frequencies. Federal Communications Commission, Washington DC, USA,
  2019. [Online]. Available:
  \url{https://docs.fcc.gov/public/attachments/FCC-20-102A1.pdf}
\BIBentrySTDinterwordspacing

\bibitem{hts1}
B.~Palacin, N.~J.~G. Fonseca, M.~Romier, R.~Contreres, J.-C. Angevain, G.~Toso,
  and C.~Mangenot, ``Multibeam antennas for very high throughput satellites in
  {Europe}: Technologies and trends,'' in \emph{Proc. {Eur.} {Conf. Antennas
  Propag.}}, Mar. 2017, pp. 2413--2417.

\bibitem{hts2}
Y.~Rahmat-Samii and A.~C. Densmore, ``Technology trends and challenges of
  antennas for satellite communication systems,'' \emph{IEEE {Trans}. Antennas
  Propag.}, vol.~63, no.~4, pp. 1191--1204, Apr. 2015.

\bibitem{hts3}
J.~M. Montero, A.~M. Ocampo, and N.~J.~G. Fonseca, ``C-band multiple beam
  antennas for communication satellites,'' \emph{IEEE Trans. Antennas Propag.},
  vol.~63, no.~4, pp. 1263--1275, Apr. 2015.

\bibitem{MUprecancel}
W.~{Zheng}, J.~{Li}, Y.~{Luo}, J.~{Chen}, and J.~{Wu}, ``Multi-user
  interference pre-cancellation for downlink signals of multi-beam satellite
  system,'' in \emph{Proc. CECNet.}, Nov. 2013, pp. 415--418.

\bibitem{LutzErich2016TtTs}
E.~{Lutz}, ``\BIBforeignlanguage{eng}{Towards the {Terabit/s} satellite -
  interference issues in the user link},'' \emph{\BIBforeignlanguage{eng}{Intl.
  J. Sat. Commun. Net.}}, vol.~34, pp. 461--482, Jun. 2015.

\bibitem{InterferenceCottatellucci}
G.~G. R.~M. L.~{Cottatellucci}, M.~{Debbah} and M.~{Neri}, ``Interference
  mitigation techniques for broadband satellite systems,'' in \emph{Proc. AIAA
  ICSSC}, vol.~1, Jun. 2006, pp. 1--13.

\bibitem{precoding}
C.~{Windpassinger}, R.~F.~H. {Fischer}, T.~{Vencel}, and J.~B. {Huber},
  ``Precoding in multiantenna and multiuser communications,'' \emph{IEEE Trans.
  Wireless Commun.}, vol.~3, no.~4, pp. 1305--1316, Jul. 2004.

\bibitem{leo_architecture}
Y.~Su, Y.~Liu, Y.~Zhou, J.~Yuan, H.~Cao, and J.~Shi, ``Broadband {LEO}
  satellite communications: Architectures and key technologies,'' \emph{IEEE
  Wireless Commun.}, vol.~26, no.~2, pp. 55--61, Apr. 2019.

\bibitem{leo_gen2}
S.~Chen, S.~Sun, and S.~Kang, ``System integration of terrestrial mobile
  communication and satellite communication —the trends, challenges and key
  technologies in {B5G} and {6G},'' \emph{China Commun.}, vol.~17, no.~12, pp.
  156--171, Dec. 2020.

\bibitem{leo_cell}
X.~Lin, S.~Cioni, G.~Charbit, N.~Chuberre, S.~Hellsten, and J.-F. Boutillon,
  ``On the path to {6G}: Embracing the next wave of low {Earth} orbit satellite
  access,'' \emph{IEEE Commun. Mag.}, vol.~59, no.~12, pp. 36--42, Dec. 2021.

\bibitem{loo}
C.~{Loo}, ``A statistical model for a land mobile satellite link,'' \emph{IEEE
  Trans. Veh. Technol.}, vol.~34, no.~3, pp. 122--127, Aug. 1985.

\bibitem{lutz}
E.~{Lutz}, D.~{Cygan}, M.~{Dippold}, F.~{Dolainsky}, and W.~{Papke}, ``The land
  mobile satellite communication channel-recording, statistics, and channel
  model,'' \emph{IEEE Trans. Veh. Technol.}, vol.~40, no.~2, pp. 375--386, May
  1991.

\bibitem{fontan}
F.~P. {Fontan}, M.~{Vazquez-Castro}, C.~E. {Cabado}, J.~P. {Garcia}, and
  E.~{Kubista}, ``Statistical modeling of the {LMS} channel,'' \emph{IEEE
  Trans. Veh. Technol.}, vol.~50, no.~6, pp. 1549--1567, Nov. 2001.

\bibitem{barts}
R.~M. {Barts} and W.~L. {Stutzman}, ``Modeling and simulation of mobile
  satellite propagation,'' \emph{IEEE Trans. Antennas Propag.}, vol.~40, no.~4,
  pp. 375--382, Apr. 1992.

\bibitem{newsimple}
A.~{Abdi}, W.~C. {Lau}, M.~{Alouini}, and M.~{Kaveh}, ``A new simple model for
  land mobile satellite channels: first- and second-order statistics,''
  \emph{IEEE Trans. Wireless Commun.}, vol.~2, no.~3, pp. 519--528, May 2003.

\bibitem{sr_mrc}
A.~M.K., ``Imperfect {CSI} based maximal ratio combining in {Shadowed-Rician}
  fading land mobile satellite channels,'' in \emph{National Conf. Commun.
  (NCC)}, Mar. 2015, pp. 1--6.

\bibitem{sr_relay}
N.~I. Miridakis, D.~D. Vergados, and A.~Michalas, ``Dual-hop communication over
  a satellite relay and {Shadowed} {Rician} channels,'' \emph{{IEEE} Trans.
  Veh. Technol.}, vol.~64, no.~9, pp. 4031--4040, Sep. 2015.

\bibitem{sr_stgem}
D.-H. Jung, J.-G. Ryu, W.-J. Byun, and J.~Choi, ``Performance analysis of
  satellite communication system under the {Shadowed-Rician} fading: A
  stochastic geometry approach,'' \emph{IEEE Trans. Commun.}, vol.~70, no.~4,
  pp. 2707--2721, Apr. 2022.

\bibitem{sumsquaredshadowrician}
G.~{Alfano} and A.~{De Maio}, ``Sum of squared {Shadowed-Rice} random variables
  and its application to communication systems performance prediction,''
  \emph{IEEE Trans. Wireless Commun.}, vol.~6, no.~10, pp. 3540--3545, Oct.
  2007.

\bibitem{closed_sum_ssr}
M.~C. Clemente and J.~F. Paris, ``Closed-form statistics for sum of squared
  {Rician} shadowed variates and its application,'' \emph{Electronics Lett.},
  vol.~50, pp. 120--121, Jan. 2014.

\bibitem{sat_ch_coraz}
G.~Corazza and F.~Vatalaro, ``A statistical model for land mobile satellite
  channels and its application to nongeostationary orbit systems,'' \emph{IEEE
  Trans. Veh. Technol.}, vol.~43, no.~3, pp. 738--742, Apr. 1994.

\bibitem{itur_681_10}
{Recommendation ITU-R P.681-10}, ``Propagation data required for the design of
  {Earth}-space land mobile telecommunication systems,'' Dec. 2017.

\bibitem{beam_coverage}
S.~Xia, Q.~Jiang, C.~Zou, and G.~Li, ``Beam coverage comparison of {LEO}
  satellite systems based on user diversification,'' \emph{IEEE Access},
  vol.~7, pp. 181\,656--181\,667, Dec. 2019.

\bibitem{leo_comparison}
I.~{del Portillo Barrios}, B.~Cameron, and E.~Crawley, ``A technical comparison
  of three low {Earth} orbit satellite constellation systems to provide global
  broadband,'' \emph{ACTA Astronautica}, vol. 159, Mar. 2019.

\bibitem{tecno_ec}
O.~B. Osoro and E.~J. Oughton, ``A techno-economic framework for satellite
  networks applied to low {Earth} orbit constellations: Assessing {Starlink},
  {OneWeb} and {Kuiper},'' \emph{IEEE Access}, vol.~9, pp. 141\,611--141\,625,
  Oct. 2021.

\bibitem{ntn_leo}
J.~{Sedin}, L.~{Feltrin}, and X.~{Lin}, ``Throughput and capacity evaluation of
  {5G} {New} {Radio} non-terrestrial networks with {LEO} satellites,'' in
  \emph{Proc. IEEE GLOBECOM}, Dec. 2020, pp. 1--6.

\bibitem{channel_measure}
C.~Loo and J.~S. Butterworth, ``Land mobile satellite channel measurements and
  modeling,'' in \emph{Proc. IEEE}, vol.~86, no.~7, Jul. 1998, pp. 1442--1463.

\bibitem{tablesofintegrals}
S.~Gradshteyn and I.~M. Ryzhik, \emph{Table of Integrals, Series, and
  Products}.\hskip 1em plus 0.5em minus 0.4em\relax New York: Academic, 2000.

\bibitem{mathhandbook}
M.~Abramowitz and I.~A. Stegun, \emph{Handbook of Mathematical
  Functions}.\hskip 1em plus 0.5em minus 0.4em\relax New York: Dover
  Publications, 1994.

\bibitem{closed_sr}
J.~F. {Paris}, ``Closed-form expressions for {Rician} shadowed cumulative
  distribution function,'' \emph{Electronics Lett.}, vol.~46, pp. 952--953,
  Jul. 2010.

\bibitem{bi_variate}
Y.~A. Brychkov and N.~Saad, ``Some formulas for the {Appell} function
  {$\mathcal{F}_1 (a, b, b'; c; w, z)$},'' \emph{Integral Transforms and
  Special Functions}, vol.~23, no.~11, pp. 793--802, Nov. 2012.

\bibitem{3gpp38811}
3GPP, ``{Technical Report} 38.811, {Study} on {New Radio (NR)} to support
  non-terrestrial networks {(NTN)},'' Jul. 2020.

\bibitem{3gpp38821}
{3GPP}, ``{Technical Report} 38.821, {Solutions} for {New Radio (NR)} to
  support non-terrestrial networks {(NTN)},'' Dec. 2019.

\bibitem{itur_618_13}
{Recommendation ITU-R P 618-13}, ``Propagation data and prediction methods
  required for the design of {Earth}-space telecommunication systems,'' Dec.
  2017.

\bibitem{iturp676}
{Recommendation ITU-R P 676-11}, ``Attenuation by atmospheric gases,'' Sep.
  2016.

\end{thebibliography}

\end{document}